\documentclass[12pt,a4paper,english]{article}
\usepackage[T1]{fontenc}
\usepackage[latin2]{inputenc}
\usepackage[english]{babel}
\usepackage{amsmath}
\usepackage{amssymb}
\usepackage{amsfonts}
\usepackage{indentfirst}
\usepackage{graphicx}

\makeatletter

\@addtoreset{equation}{section}
\makeatother

\newcommand{\bea}{\begin{eqnarray}}
\newcommand{\eea}{\end{eqnarray}}
\newcommand{\be}{\begin{equation}}
\newcommand{\ee}{\end{equation}}

\def\beq#1\eeq{\begin{align}#1\end{align}}

\def\Tr{{\rm Tr}}

\def\G{\Gamma}

\newcommand{\bC}{\ensuremath{\mathbb{C}}}

\newcommand{\bP}{\ensuremath{\mathbb{P}}}

\newcommand{\bZ}{\ensuremath{\mathbb{Z}}}

\newcommand{\pg}{\overset{+}{\succ}}
\newcommand{\pl}{\overset{+}{\prec}}

\begin{document}

\begin{titlepage}
\begin{flushright}
\normalsize
 AEI-2010-091\\ 
 CALT-68-2786\\
 IPMU10-0078

\bigskip

\end{flushright}

\begin{center}
\LARGE{Wall Crossing As Seen By Matrix Models}
\end{center}

\vfil
\bigskip

\begin{center}

Hirosi Ooguri$^{a,b,c}$, Piotr Su{\l}kowski$^{a,}$\footnote{On leave from University of Amsterdam and So{\l}tan Institute for Nuclear Studies, Poland.}, Masahito Yamazaki$^{b}$

\bigskip
\bigskip
\itshape

$^a$ California Institute of Technology, Pasadena CA 91125, USA

\medskip
$^b$ Institute for the Physics and Mathematics of the Universe,\\ 
 University of Tokyo, Kashiwa 277-852, Japan

\medskip
$^c$ Max-Planck-Institut f\"ur Gravitationsphysik, 
D-14476 Potsdam, Germany

\end{center}

\vfil
\bigskip

\begin{center}
{\bfseries Abstract}
\end{center}
The number of BPS bound states of D-branes on a Calabi-Yau manifold 
depends on two sets of data, the BPS charges and the stability conditions.
For D0 and D2-branes bound to a single D6-brane wrapping a Calabi-Yau 3-fold
$X$, both are naturally related to the K\"ahler moduli space ${\cal M}(X)$.
We construct unitary one-matrix models which count such BPS states
for a class of toric Calabi-Yau manifolds at infinite 't Hooft coupling. 
The matrix model for 
the BPS counting on $X$ turns out to give the topological 
string partition function for another Calabi-Yau manifold $Y$, 
whose K\"ahler moduli space ${\cal M}(Y)$ contains two copies 
of ${\cal M}(X)$, one related to the BPS charges and another 
to the stability conditions. The two sets of data are
unified in ${\cal M}(Y)$. The matrix models have 
a number of other interesting features. 
They compute spectral curves and mirror maps
relevant to the remodeling conjecture.
For finite 't Hooft coupling
they give rise to yet more general geometry 
$\widetilde{Y}$ containing $Y$.

\bigskip

\vfill

\end{titlepage}

\section{Introduction}

The topological string theory has deep connections to 
a variety of BPS counting problems 
in string theory \cite{GV,OSV}. In this paper, we focus 
on the generalized Donaldson-Thomas (DT) invariants, namely 
the numbers of D0 and D2 bound states on a single
D6 brane wrapping a Calabi-Yau 3-fold $X$. The DT invariants
are background dependent. As we vary the K\"ahler moduli
of $X$ and cross a wall of marginal stability, the
numbers can jump. To count BPS bound states, we have to
specify the stability condition, $i.e.$ the chamber in the moduli space
 where we perform the counting. 
%Each chamber is surrounded by walls of
%marginal stability and its location corresponds to
%the choice of stability conditions to define bound states. 
Thus, the DT invariant depends on two
sets of data, the BPS charges and the stability conditions. 
In particular, the commutative DT invariants are defined in 
the chamber corresponding to the infinity in the K\"ahler
moduli space, while the non-commutative DT invariants
are defined in the chamber containing the origin. 

It is convenient to introduce the generating function $Z_{{\rm BPS}}$ 
of the DT invariants $\Omega_{\alpha, \beta}(n)$,
\be Z_{{\rm BPS}}(q, Q; n) = \sum_{\alpha, \beta} 
\Omega_{\alpha, \beta}(n) q^\alpha Q^\beta, 
\label{BPSpartition}
\ee
where $\alpha \in \mathbb{Z}$ is the D0 brane charge, $\beta \in H_2(X, \mathbb{Z})$
are the D2 brane charges, and $n$ is a set of parameters which specify
the chamber in the K\"ahler moduli space. In this paper we consider toric Calabi-Yau manifolds without compact
4-cycles, see figure \ref{fig-strip}. For a manifold $X$ in this class,
it was shown in \cite{AOVY} that $Z_{{\rm BPS}}$ 
is given by a certain reduction of the square
of the topological string partition function $Z_{{\rm top}}(q, Q)$,
\be Z_{{\rm BPS}}(q, Q; n)  
= Z_{{\rm top}}(q, Q) \cdot Z_{{\rm top}}(q, Q^{-1})\Big|_{{\rm reduction~at}~n}.
\label{AOVY}
\ee
In this case, 
$Z_{{\rm top}}(q,Q)$ is expressed as a product in the
harmonic oscillator form. The reduction means dropping 
an appropriate set of harmonic oscillator factors 
from $|Z_{{\rm top}}|^2$ corresponding 
to D0/D2 states that do not bind with the single D6 brane
in the chamber $n$.

Both $Q$ and $n$ are related to the K\"ahler moduli space
${\cal M}(X)$ of $X$. The relation of $n$ to the moduli space is
clear since it specifies a chamber in ${\cal M}(X)$. 
It is also natural to identify $Q=e^{-t}$ in (\ref{BPSpartition})
with $t$ being flat coordinates of ${\cal M}(X)$ since the BPS charges
couple to the areas of the corresponding homology cycles,
$i.e.$ the K\"ahler moduli. However, these two data appear 
asymmetrically in (\ref{AOVY}). 
In this paper, we will present another connection of 
$Z_{{\rm BPS}}$ to the topological string theory, in which they are
treated more symmetrically. We will show that there is another 
Calabi-Yau manifold $Y$, whose K\"ahler moduli space
${\cal M}(Y)$ contains two copies of ${\cal M}(X)$, and the topological
string partition function for $Y$ is related to 
$Z_{{\rm BPS}}$ for $X$. For example, when $X$ is the resolved conifold with 
${\rm dim}_{\bC}\ {\cal M}(X)=1$, the corresponding $Y$ is the suspended
pinch point (SPP) geometry with ${\rm dim}_{\bC}\ {\cal M}(Y)=2$.
Similarly, when $X$ is $\mathbb{C}^3/\mathbb{Z}_2$, the corresponding
$Y$ is $\mathbb{C}^3/\mathbb{Z}_3$.

We will find this relation by constructing the unitary one-matrix 
model whose partition function $Z_{{\rm matrix}}(q,Q; n)$ is related to 
$Z_{{\rm BPS}}(q,Q;n)$. In particular, $Z_{{\rm matrix}}$ is equal to 
$Z_{{\rm BPS}}$ in the non-commutative chamber ($n=0$) and is equal to
$Z_{{\rm top}}(X)$ in the commutative chamber ($n=\infty$).   
To derive the matrix model, we start with the crystal melting 
model \cite{ORV,OY1} to count the generalized DT invariants,
and use the vertex operator formalism \cite{2009pyramid,NagaoVO},
in which the partition function is expressed 
as correlators of exponentials 
of fermion bilinears. The correlators are defined for 
all chambers in the K\"ahler moduli space, and 
we can transform the computation into unitary matrix integrals. 
This construction is closely connected to the free fermion picture 
for the topological string and Seiberg-Witten theory 
developed in \cite{ADKMV,dhsv,dhs}.
Equivalently, we can also express the partition 
function as a sum over non-intersecting 
paths following and generalizing \cite{EynardTASEP}, which 
gives yet another derivation of such matrix models. 

One interesting feature of our matrix model for the conifold, in the commutative chamber,
 is its close relation to the so-called Chern-Simons matrix model 
of \cite{mm-lens,CSmatrix}. In the commutative chamber in our model, $Q$ is the only parameter,
 and it appears only in the potential.
The Chern-Simons matrix model also depends on a single parameter, which is the 't Hooft coupling.
It turns out that these two parameters play the same role in the partition function in both models. 
Moreover one can consider a model with non-zero values of both these parameters. 
From this viewpoint, 
departure from the commutative 
to arbitrary chamber can be interpreted as turning on yet another parameter. 
In general one can consider simultaneously non-zero values of all three parameters:
$Q$, chamber dependence, and 't Hooft coupling. 
This gives rise to the spectral curve encoding yet 
more general Calabi-Yau manifold  $\widetilde{Y}$, 
which contains the manifold $Y$ described above. 
When $X$ is the conifold, 
$\widetilde{Y}$ is a symmetric resolution of $\mathbb{C}^3/\mathbb{Z}_2
\times \mathbb{Z}_2$, while $Y$ is the SPP geometry as we mentioned
in the above. Such $\widetilde{Y}$ can in principle be constructed 
for any initial
toric manifold $X$.%, and we find its combinatorial interpretation on the level 
%of a crystal melting model: finite 't Hooft coupling $g_S N$ corresponds 
%to an insertion of a wall at location $N$ in a corresponding crystal.

As a bonus of our matrix model construction, 
it sheds new light on the remodeling conjecture.
It has been conjectured in \cite{BKMP} that the topological string partition
function for this class of Calabi-Yau manifolds is completely characterized
by the recursion relations of \cite{EO}, applied to the curve
which should be identified with the mirror curve of a given manifold. 
Such recursion relations would arise if we had a matrix model formulation 
of the topological strings. In this paper we provide a construction of such matrix models
in several instructive cases,
and verify that to the leading order their spectral curves agree with relevant mirror curves,
which is an important step towards a proof of the remodeling conjecture.
%In case of $\bC^3$, the conifold and $\bC^3/\mathbb{Z}_2$, 
%our direct derivation proves the remodeling proposal. 
%
%
%we show that by construction
%partition functions of our matrix models agree with topological string partition functions, and
%for $\bC^3$ and the conifold we explicitly verify that their spectral curves agree with mirror curves.
We expect that application of our methods 
should lead to analogous results in the general case of toric manifold without
compact 4-cycles.

%While the recursion relations of \cite{EO} in principle arise from matrix model construction,
%nd should be applied to the matrix model spectral curve, the 
%onjecture of \cite{BKMP} does not provide a hint on how to construct 
%matrix models relevant for the B-model, or whether 
%hey exist at all. Matrix models for 
%PS counting which we derive here have partition functions which reduce to the topological string 
%artition function in the commutative chamber, and they are natural candidate for a microscopic
%ompletion of the remodeling conjecture. For $\bC^3$
%and the conifold, we prove that spectral curves of our matrix models indeed 
%gree with appropriate mirror curves. 
%ven though we do not apply recursion relations to those curve (which has already been
%one in \cite{BKMP}), our direct derivation of mirror curves proves the
%emodeling proposal. It is straightforward to generalize this approach 
%o arbitrary manifold in the class we consider in this paper.
%part from a subtlety related to certain prefactors in generic chambers, 
%ur approach generalizes the remodeling conjecture to arbitrary chambers.

We also note that, for the case of $\bC^3$, a similar approach was presented 
in \cite{EynardTASEP}. For an earlier related work, see \cite{Okuda}. 
Matrix models for other Calabi-Yau manifolds in the commutative chamber 
were derived from
the topological vertex formalism or Nekrasov partition functions 
in \cite{eynard-planch,SW-matrix,matrix2star,2009betaMatrix}.
In the course of this work we received the paper 
\cite{EynardTopological}, 
in which matrix models are derived in the commutative chamber also from 
the topological vertex perspective.
Related ideas have been considered in \cite{DSV,mina}.

This paper is organized as follows. In section 2 we introduce
matrix models for BPS counting and explain how they are related to the DT invariants. 
In section 3 we examine spectral curves of the matrix models
and identify the corresponding Calabi-Yau geometries.
In particular, when $X$ is the resolved conifold, 
we also identify the total geometry $\widetilde{Y}$ 
for finite 't Hooft coupling, and discuss its relations to
the Chern-Simons matrix model. 
The derivation of the matrix model is given in section 4. 
We end with summary and discussion on future research directions in
section 5.

\section{Matrix Models}

In this section, we will present matrix models which count the
DT invariants, namely the number of BPS states of D0 and D2-branes 
bound to a single D6 wrapping a Calabi-Yau manifold $X$. 
In general these are matrix models for unitary matrices of infinite size, 
and arise from crystal melting interpretation of BPS generating
functions. The derivation of these matrix models will be given in section 4. 

\subsection{$\mathbb{C}^3$}    \label{ssec-C3matrix}

When $X=\mathbb{C}^3$, the generating function of BPS invariants is given
by the MacMahon function which counts plane partitions. We find that this BPS generating function is equal
to the partition function of the matrix model given by
\be
 Z_{{\rm matrix}}(q) = \int dU \ {\rm det}\ \Theta(U| q),
\label{CthreeMatrix}
\ee
where the integral is over the unitary group $U(N)$, and we are
interested in the limit of $N\rightarrow \infty$. The integrand 
is given by the theta-product,
\be 
\Theta(u| q) = \prod_{k=0}^\infty (1 + u q^k)(1+u^{-1}q^{k+1}).
\ee
To perform the integral (\ref{CthreeMatrix}), it is convenient to 
diagonalize $U = {\rm diag}(u_1, ..., u_n)$ and 
to consider the integral over eigenvalues $u_i = e^{i\phi_i}$ as
\be 
   Z_{{\rm matrix}}(q) = \int  
    \prod_i d\phi_i \ \Theta(e^{i\phi_i}|q)\ 
\prod_{i<j} (e^{i\phi_i} - e^{i\phi_j})
              (e^{-i\phi_i} - e^{-i\phi_j}) .
\label{ThetaIntegrand}
\ee
As usual, the two factors of the Vandermonde determinant 
come from the integral over off-diagonal elements of $U$.
To perform the integral (\ref{ThetaIntegrand}) over
eigenvalues, 
we expand the integrand in powers of $q$, 
\begin{equation*}
   \Theta(e^{i\phi}| q) = 1 + e^{i\phi} 
+ (1+ e^{-i\phi}+  e^{i\phi} + e^{2i\phi})\ q + 
(2+ e^{-i\phi}+2e^{i\phi}+e^{2i\phi})\ q^2
+ \cdots,
\end{equation*}
and pick up appropriate combinations of $e^{\pm i \phi_i}$'s from the
measure factor in (\ref{ThetaIntegrand}) 
to cancel the $\phi$-dependence in $\Theta(e^{i\phi}|q)$.
In this way, we can directly verify that the integral 
gives the MacMahon function, 
\be
Z_{{\rm matrix}}(q) = 1 + q + 3 q^2 + 6 q^3 + 13q^4 + \cdots = 
\prod_{k=1}^\infty\frac{1}{ (1-q^k)^k}.
\label{CthreeMc}
\ee
This is indeed the generating function of plane partitions and 
reproduces the counting 
of the DT invariants on $\mathbb{C}^3$
if we identify the power of $q$ as the D0 brane charge.
In this case, there is no distinction between 
commutative and non-commutative chambers. 

To relate this to the Chern-Simons matrix model, we make the
identification of $q=e^{-g_s}$, where $g_s$ is the string coupling constant. 
For small $g_s$, the modular transformation of $\Theta$
with respect to $g_s$ gives
\be
 \Theta(e^{i\phi}| e^{-g_s}) = 
e^{-\frac{\phi^2}{2g_s}}\cdot \left(1  +  O(e^{-\frac{1}{g_s}})\right).   
\ee   
If we ignore non-perturbative terms in $g_s$, this is equal to the
integrand for the unitary Gaussian matrix model derived from the 
Chern-Simons theory on the conifold \cite{CSmatrix}.
In fact, (\ref{CthreeMatrix}) itself has also been proposed for 
the topological string theory on the conifold in \cite{Okuda}, 
whose approach is a special case of our fermionic derivation applied to $\bC^3$
as we will see below. 
The K\"ahler moduli $T$ of the resolved conifold is given by the 't Hooft
coupling,
\be
 T = g_s N.
\ee

We are interested in the $N\rightarrow \infty$ limit for fixed $g_s$, namely
$T\rightarrow \infty$. It is shown in \cite{Okuda} that the model (\ref{ThetaIntegrand})
with finite $N$ has an interpretation of counting plane partitions in a container with
a wall at position $N$. As we will
discuss in the next section, finite 't Hooft parameter
has similar wall interpretation in our more general models.
From this perspective, $N\to\infty$ limit in $\bC^3$ model corresponds to 
computing all plane partitions. This limit suppresses instanton corrections on the conifold,
leaving only contributions from constant maps. For a general Calabi-Yau
manifold, the sum over constant maps gives the MacMahon function to the power 
of $\chi/2$, where $\chi$ is the Euler characteristics of the Calabi-Yau 
manifold. Since $\chi=2$ for the resolved conifold, we find that the
$N\rightarrow \infty$ limit gives one power of the MacMahon function,
reproducing (\ref{CthreeMc}).

\subsection{Conifold}

The K\"ahler moduli space of the resolved conifold is 
complex 1-dimensional, and it is divided into chambers
parametrized by an integer $n$, which is the 
integer part of the B-field flux through the $\bP^1$ \cite{AOVY}. 
The non-commutative chamber corresponds to $n=0$ and the
commutative chamber is at $n=\infty$. 

We find the following matrix model in the non-commutative chamber:
\be
  Z_{{\rm matrix}}(q, Q; n=0) 
= \int dU\ {\rm det}\left( \frac{ \Theta(U| q)}{\Theta(QU| q)}\right),
\ee
where
\be Q = e^{-t} \ee
keeps track of the D2 brane charge. By expanding the integrand 
in powers of $q$ and
by performing the integral over $U(N)$ in the $N \rightarrow \infty$
limit as in the previous
example, we can verify that
\begin{align}
  Z_{{\rm matrix}}(q,Q; n=0)
&= 1 + (2 - Q^{-1}-Q)q + (8-4Q^{-1}-4Q) q^2 + \cdots \nonumber \\
&= \prod_{k=1}^\infty \frac{(1-Q q^k)^k (1-Q^{-1}q^k)^k}{(1-q^k)^{2k}}.
\end{align}
This reproduces $Z_{{\rm BPS}}(q,Q;n=0)$ in the non-commutative chamber. 

For a general chamber, the BPS partition
function is given by
\be
 Z_{{\rm BPS}}(q,Q;n) = 
\prod_{k=1}^\infty \frac{(1-Q q^k)^k (1-Q^{-1}q^{n+k})^{n+k}}{(1-q^k)^{2k}}.    \label{ZBPS-n-conifold}
\ee
The free fermion expression for $Z_{{\rm BPS}}(q,Q; n)$, 
discussed in section 4, gives rise to the following
matrix integral,
\be
 Z_{{\rm matrix}}(q,Q;n) = 
\int dU\ {\rm det}\left(\frac{\Theta(U|q)}{\Theta(QU|q)}
  \prod_{k=1}^{n}(1+Q^{-1}U^{-1}q^k)\right).
\label{conifoldmatrixone}
\ee
The BPS partition function and the matrix model partition function are
related as 
\be 
Z_{{\rm BPS}}(q,Q;n) = C_n \cdot Z_{{\rm matrix}}(q,Q;n),   \label{ZBPS-Zmatrix}
\ee
where the prefactor $C_n$ is given by
\be
   C_n =\prod_{k=1}^n  \frac{1}{(1-q^k)^k}
\prod_{k=n+1}^\infty \left( \frac{1-Q^{-1}q^{k}}{1-q^k}
\right)^n.
\label{Cnconifold}
\ee
We also verfied (\ref{ZBPS-Zmatrix}) by expanding matrix model integrand 
and integrating it term by term. 
The origin of the prefactor $C_n$ will be explained in section 4. 
Note that this prefactor is trivial in the non-commutative chamber, 
$C_{n=0}=1$.

It is known that %for a toric Calabi-Yau manifold without compact 4-cycles, 
the BPS partition function in the commutative chamber and the
topological string partition function are identical, up to one power
of the MacMahon function,
\be
 Z_{{\rm BPS}}(q,Q; n=\infty) = 
Z_{{\rm top}}(q=e^{-g_s},Q=e^{-t})\cdot \prod_{k=1}^\infty\frac{1}{ (1-q^k)^k}.
\ee
Since the prefactor $C_n$ reduces to the MacMahon function in the
commutative limit,
\be
C_{n=\infty} =  \prod_{k=1}^\infty \frac{1}{(1-q^k)^k},
\ee
the matrix model partition function 
gives precisely the topological string partition function in the
commutative chamber,
\begin{align}
Z_{{\rm matrix}}^{(n=\infty)} & = \int dU \ {\rm det}\left(
\prod_{k=0}^\infty\frac{(1+U q^k)(1+U^{-1}q^{k+1})}{(1 + Q U q^k)}\right) \nonumber \\
&= Z_{{\rm top}}(q,Q).
\end{align}
In this way, the matrix model partition function 
$Z_{{\rm matrix}}(q,Q;n)$ interpolates between 
$Z_{{\rm BPS}}$
in the non-commutative chamber and 
$Z_{{\rm top}}$ in the commutative chamber. 

\subsection{$\mathbb{C}^3/\mathbb{Z}_2$}   \label{ssec-C3Z2}

Another toric Calabi-Yau manifold with 
${\rm dim}_{\bC}\ {\cal M}(X)=1$ is $\mathbb{C}^3/\mathbb{Z}_2$.
The matrix model for the non-commutative chamber is given by
\be
 Z_{{\rm matrix}}(q,Q;n=0) 
= \int dU\ {\rm det}\left( \Theta(U|q) \Theta(QU| q) \right).
\ee
For a general chamber, we can write the explicit product form
of the BPS generating function as a matrix integral
\bea
 Z_{{\rm BPS}}(q,Q;n) & =  & \prod_{k=1}^\infty (1-q^k)^{-2k}(1-Q q^k)^{-k} (1-Q^{-1}q^{n+k})^{-n-k} = \nonumber \\
& = & C_n \cdot \int dU \
{\rm det}\left( \frac{\Theta(U|q) \Theta(QU| q)}{
\prod_{k=1}^{n}(1+Q^{-1}U^{-1}q^k)}\right),
\label{matrixorbifold}
\eea
with the prefactor
\be
C_n = \prod_{k=1}^n \frac{1}{(1-q^k)^k} 
\prod_{k=n+1}^\infty \left( \frac{1}{(1-q^k)(1-Q^{-1}q^k)} \right)^n.
\label{CnZ2}
\ee
This can be verified explicitly by expanding both sides of 
(\ref{matrixorbifold}) in powers of $q$. 

Again, in this case, we have $C_{n=0}=1$ and 
$C_{n=\infty} =\prod_k (1-q^k)^{-k}$. Thus, the matrix model 
partition function interpolates between $Z_{{\rm BPS}}$ in the
non-commutative chamber and $Z_{{\rm top}}$ in the commutative chamber,
\begin{align}
\begin{split}
& Z_{{\rm matrix}}(q,Q; n=0) =
Z_{{\rm top}}(q,Q)\cdot Z_{{\rm top}}(q,Q^{-1}) = Z_{{\rm BPS}}(q, Q; n=0), 
\\
& Z_{{\rm matrix}}(q,Q; n=\infty) = Z_{{\rm top}}(q,Q).
\end{split}
\end{align}

\subsection{General Toric Calabi-Yau Manifold}    \label{subsec.generalmatrix}

A toric Calabi-Yau 3-fold $X$ without compact 4-cycle consists of 
a chain of $\mathbb{P}^1$'s, which is resolved either 
by ${\cal O}(-1,-1)$
or ${\cal O}(-2,0)$. The topological string partition function
for such a Calabi-Yau manifold is given by \cite{AKMV,IK}
\be
Z_{{\rm top}}(q,Q)=
\left(\prod_{k=1}^\infty \frac{1}{(1-q^k)^k}\right)^{\chi/2}
\prod_{1\leq i < j \leq \chi-1} \prod_{k=1}^\infty 
(1 - Q_i \cdots Q_j q^k)^{s_i \cdots s_j k},
\ee
where $\chi$ is the Euler characteristics of $X$,
the number of $\mathbb{P}^1$'s is $(\chi-1)$, and 
$Q_1,...,Q_{\chi-1}$
are the K\"ahler moduli that measure their sizes.
Depending on whether the $i$-th $\mathbb{P}^1$ is resolved 
by ${\cal O}(-1,-1)$ or ${\cal O}(-2,0)$, we set $s_i=-1$ or $+1$.

The BPS partition function in
the non-commutative chamber is given by
\be
 Z_{{\rm BPS}}(q,Q;n=0) = Z_{{\rm top}}(q, Q)\cdot Z_{{\rm top}}(q, Q^{-1}).
\ee
This is reproduced by the matrix model partition function,
\be
Z_{{\rm matrix}}(q,Q;n=0) 
= \int dU \det \ \prod_{i=1}^{\chi-1} 
\Theta(s_1Q_1 \cdots s_i Q_i U| q)^{s_1 \cdots s_i}.
\ee
Following the procedure described in section 4, it is possible 
to write down matrix models for 
other chambers. However, we have not attempted to
derive a closed-form expression of the matrix model potential
for a general chamber.

\section{Spectral Curves and Geometric Unification}

The eigenvalue distribution of the large $N$ matrix model is
controlled by a spectral curve. In particular, the resolvent is 
a one-form on the curve and the large $N$ effective action is
evaluated by its period integral on the curve. It has been argued 
from several viewpoints that spectral curves of matrix models
arising from the topological string theory should be
related to the geometry of the corresponding Calabi-Yau manifold \cite{BKMP,SW-matrix,DV}. 

In this section, we will identify the spectral curves and
the corresponding Calabi-Yau geometries $Y$ for the
matrix models defined in the previous section. These geometries arise in the 
limit of infinite 't Hooft coupling.
In a nontrivial case of $X\neq \mathbb{C}^3$,
they contain two copies of the initial Calabi-Yau manifold $X$
 for a generic chamber.
For the conifold case, we will
analyze in detail yet more general geometry 
$\widetilde{Y}$ which arises for finite 
't Hooft coupling, as well as reveal close relation 
between conifold matrix model
in the commutative chamber and so-called Chern-Simons 
matrix model \cite{mm-lens,CSmatrix}. 

\subsection{$\mathbb{C}^3$}

As a warm-up exercise, let us describe
the unitary Gaussian model, discussed in
\cite{CSmatrix,Okuda,Marino}, 
\be
   Z_{{\rm matrix}}(q) = \int  
    \prod_i d\phi_i  
\ e^{-\frac{1}{2g_s}\phi_i^2}
\ \prod_{i<j} (e^{i\phi_i} - e^{i\phi_j})
              (e^{-i\phi_i} - e^{-i\phi_j}) .    \label{Gaussian-C3}
\ee
Since $\phi_i$'s are periodic variables, it may appear
unnatural to have the non-periodic potential, $e^{-\frac{1}{2g_s}\phi_i^2}
$. In our construction, it is the $g_s \rightarrow 0$ limit of
the periodic integrand given in (\ref{ThetaIntegrand}). 
The integrand $\Theta(e^{i\phi}|q)$ has a series of
zeros at $\phi = i k g_s$ with $k \in \mathbb{Z}$, which becomes a branch
cut along the imaginary axis in the limit $g_s \rightarrow 0$. 

The spectral curve for the unitary Gaussian matrix model 
is given by the equation \cite{CSmatrix,Marino}
\be
  e^x + e^y + e^{x-y - T} + 1 = 0, 
\ee
where $T=Ng_s$ is the 't Hooft coupling. The corresponding
Calabi-Yau manifold is the mirror of the resolved conifold. 
In the limit of $T\rightarrow \infty$, the curve reduces to
\be
  e^x + e^y + 1 = 0, 
\ee
which is the mirror of $\mathbb{C}^3$. This result also arises
as a special case $Q,e^{-T},\mu\to 0$ of a derivation of the conifold curve
presented in the next section.

\subsection{Conifold}

In this section we analyze conifold matrix model in all chambers. 
From the form of the spectral curve and for finite 't Hooft coupling
 we identify the total manifold $\widetilde{Y}$ to be a resolution of
the $\mathbb{C}^3/\mathbb{Z}_2 \times \mathbb{Z}_2$ orbifold. In the 
limit of infinite  't Hooft coupling, $\widetilde{Y}$ reduces to 
the susspended pinch point (SPP) geometry $Y$,
 which contains two copies of the initial conifold geometry.

To examine the $g_s \rightarrow 0$ limit of the matrix model 
for the conifold, let us look at the integrand of (\ref{conifoldmatrixone}).
We find it convenient to choose the freedom of renaming $U\to U^{-1}$ 
described in section \ref{ssec-MatrixFermions} and consider an equivalent integrand
\be
\frac{\Theta(U^{-1}|q)}{\Theta(QU^{-1}|q)}
  \prod_{k=1}^n(1+Q^{-1}U q^k)
 = \prod_{k=0}^\infty 
\frac{(1+ U^{-1} q^k)(1+U q^{k+1})}{(1 + QU^{-1} q^k)(1+ Q^{-1}Ue^{-\tau}q^{k+1})} .
\label{conifoldintegrand}
\ee
Here and in what follows we set 
$$
\tau = n g_s.
$$ 
In order to retain interesting dependence on the
chamber parameter $n$, we should take the limit 
$g_s \rightarrow 0$ in such a way that $\tau$ is held finite. 
By using the identity,
\be
\log \ \prod_{k=1}^\infty (1 + U q^k)
\sim - \frac{1}{g_s}{\rm Li}_2(-U), ~~ (|g_s| \ll 1),
\ee
where ${\rm Li}_2$ is the dilogarithm function, 
the integrand (\ref{conifoldintegrand}) can be approximated by 
$e^{-\frac{1}{g_s} V(U)}$ with
\be
 V(U) = T\log U + {\rm Li}_2(-U) + {\rm Li}_2(-U^{-1})- {\rm Li}_2(- QU^{-1})
        - {\rm Li}_2(-Q^{-1}e^{-\tau}U),     \label{Vconifold}
\ee
where we also took into account the shift (\ref{measure-T}) of the potential
which arises from the transformation of the measure to the form which
includes the Vandermonde determinant. Therefore
\be
\partial_U V = \frac{T-\log (U+Q) + \log\big(1 + \frac{U}{Q e^{\tau}}\big)}{U}.    
\ee

\begin{figure}[htb]
\begin{center}
\includegraphics[width=0.35\textwidth]{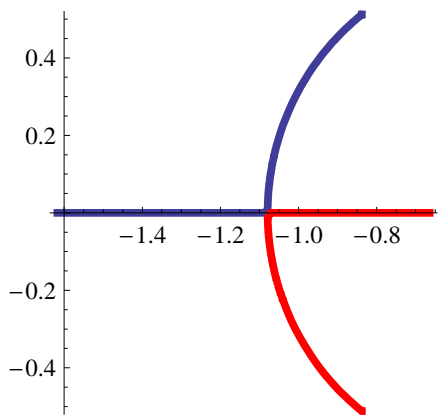} 
\begin{quote}
\caption{\emph{Behavior of cut end-points $a_+$ (in blue) and $a_-$ (in red) given in (\ref{aAll}), for fixed $\epsilon,Q$ and varying $\mu$. For $\mu<1$, end-points $a_{\pm}$ are complex conjugate to each other. For $\mu=1$ we have $a_+=a_-=-1-\frac{\epsilon^2(1-Q)}{1-\epsilon^2}$ and the cut shrinks to zero size. For $\mu>1$ both $a_{\pm}$ are real and spread in opposite directions.}} \label{fig-cutends}
\end{quote}
\end{center}
\end{figure}

Let us define the resolvent $\omega(u)$ by
\be
  \omega(u) = \frac{1}{N} \left\langle {\rm tr} \left( \frac{1}{u - U}\right) \right\rangle,
\ee
where the expectation value is taken over the large $N$ eigenvalue
distribution of $U$. As we expect to find a genus 0 curve, we postulate 
the existence of a one-cut solution. In this case, in the weakly coupled phase
of a unitary matrix model, 
the resolvent can be computed using the standard 
Migdal integral %applied to the unitary matrix model
\be
\omega(u) = \frac{1}{2T} \oint \frac{d v}{2\pi i} \frac{\partial_v V(v)}{u-v}\frac{\sqrt{(u-a_+)(u-a_-)}}{\sqrt{(v-a_+)(v-a_-)}},  \label{MigdalTxt}
\ee
where the integration contour encircles counter-clockwise the endpoints of the cut $a_{\pm}$. 
We perform this computation in appendix A and find
\be
\omega(u) = \frac{1}{uT} 
\log \left( \frac{\sqrt{(a_++Q)(a_--u)}-\sqrt{(a_-+Q)(a_+-u)}}
{\sqrt{(a_++Qe^{\tau})(a_--u)}-\sqrt{(a_-+Qe^{\tau})(a_+-u)}}  
\frac{u+Qe^{\tau}}{u+Q}    
\frac{e^{T/2}}{Q^{1/2} e^{\tau/2}} \right).
\label{conifoldresolvent}
\ee
%where $a_+$ and $a_-$ are end-points of the branch cut. 
This form of the resolvent already takes into account the boundary condition 
\be
 \omega(u\rightarrow \infty) \sim \frac{1}{u}.    \label{eq.omegaboundary}
\ee
This condition also gives rise to two equations on the location of $a_{\pm}$
\bea
\frac{\sqrt{a_{+}+Q}-\sqrt{a_{-}+Q}}{\sqrt{a_{+}+Qe^{\tau}}-\sqrt{a_{-}+Qe^{\tau}}} & = &  Q^{\frac{1}{2}} e^{(\tau+T)/2} ,     \label{ab-eq1}  \\
\frac{\sqrt{(a_{+}+Q)a_{-}}-\sqrt{(a_{-}+Q)a_{+}}}{\sqrt{(a_{+}+Qe^{\tau})a_{-}}-\sqrt{(a_{-}+Qe^{\tau})a_{+}}} & = & Q^{\frac{1}{2}}  e^{-(\tau+T)/2} .    \label{ab-eq2}
\eea
With some effort these equations can be solved in the exact form
\begin{align}
\begin{split}
a_\pm & =  -1 + \epsilon^2\ \frac{(1-\mu)(1-\mu\epsilon^2) + 
(1-Q)(1+\mu\epsilon^2 - 2\mu)}{(1-\mu\epsilon^2)^2} 
 \label{aAll} \\
& ~~~~\pm 2 i \epsilon\ \frac{\sqrt{(1-Q)(1-\epsilon^2)(1-\mu)
(1-Q\mu\epsilon^2)}}{(1-\mu\epsilon^2)^2},
\end{split}
\end{align}
see figure \ref{fig-cutends}. The parameters $\mu$ and $\epsilon$ are related to the chamber
number $\tau$ and the 't Hooft parameter $T$ as
\be 
  \mu = Q^{-1} e^{-\tau},~~ \epsilon=e^{-T/2}.
\ee
From the form of (\ref{aAll}) it is clear that in the saddle point approximation 
the cuts are deformed and do not lay on a unit circle, 
but (as often happens in similar situations) on arcs which are deformations thereof. 
One can also verify that $\omega(u)$ given by (\ref{conifoldresolvent}) is
indeed a solution to the Riemann-Hilbert problem,
\be
 \omega_+(u) + \omega_-(u) = \frac{1}{T} \frac{\partial V }{\partial u},
\ee
where $\omega_\pm$  are the values of $\omega(u)$ 
right above and below the branch cut. 
From the resolvent one can also find the eigenvalue density (for $\mu<1$)
$$
\rho(u) = \omega_+(u) - \omega_-(u) = \frac{1}{uT} \log\left( \frac{ (1+\mu\epsilon^2)u + 1 +Q\epsilon^2 - (1-\mu \epsilon^2) \sqrt{(u-a_+)(u-a_-)}}{ (1+\mu\epsilon^2)u + 1+Q\epsilon^2 + (1-\mu \epsilon^2)\sqrt{(u-a_+)(u-a_-)}}  \right).
$$

To identify the spectral curve we note first that the non-trivial part of the resolvent
takes the form (for $\mu<1$)
\be
\omega(u) \sim  % & = & \frac{1}{pT} \log \Big(\frac{e^{T/2}}{Q^{1/2} e^{\tau/2}}
% \frac{\sqrt{(a+Q)(a+Qe^{\tau})}-\sqrt{(b+Q)(b+Qe^{\tau})}}{(p+Q)(b-a)}  \Big) + \nonumber \\ 
  \frac{1}{uT} \log \Big( -u - \frac{1+Q\epsilon^2}{1+\mu\epsilon^2} + \frac{1-\mu \epsilon^2}{1+\mu\epsilon^2} \sqrt{(u-a_+)(u-a_-)}\Big)  .    \label{omega-coni}
\ee
After identification $x=uT\, \omega(u)$, and setting $u=e^y$, we find that $e^x$ and $e^y$ satisfy
a polynomial equation. Appropriate constant shifts of $x$ and $y$ transform this equation 
into the following form
\be
  e^{x+y} + e^x + e^y 
+ Q_1\ e^{2x}
 + Q_2\ e^{2y} + Q_3= 0,
\label{symmetricresolution}
\ee   
where
\begin{align}
\begin{split}
& Q_1 = \epsilon^2 \cdot 
\frac{1+\mu Q}{(1+\mu \epsilon^2)(1+Q\epsilon^2)}, \\
& Q_2 = \mu \cdot
 \frac{1+Q\epsilon^2}{(1+\mu Q)(1+\mu\epsilon^2 )}, \\
& Q_3 = Q \cdot \frac{1+\mu \epsilon^2}{(1+\epsilon^2 Q)(1+\mu Q)}.
\end{split}
\end{align}
The above equation represents the spectral curve we have been after.
It is interesting that the curve (\ref{symmetricresolution}) 
is symmetric under exchanges of $Q$, $\mu = Q^{-1}q^n$ and $\epsilon^2 
= e^{-T}$. Namely, the original K\"ahler moduli $Q$ 
of the resolved conifold, the chamber parameter $n$ and the 't Hooft
parameter $T$ appear symmetrically in the spectral curve. 
We also note that the above form of the curve, 
as well as the density and the resolvent
given in (\ref{omega-coni}), are valid for $|\mu| < 1$. For $|\mu|> 1$, an
appropriate analytic continuation is required.

\begin{figure}[htb]
\begin{center}
\includegraphics[width=0.7\textwidth]{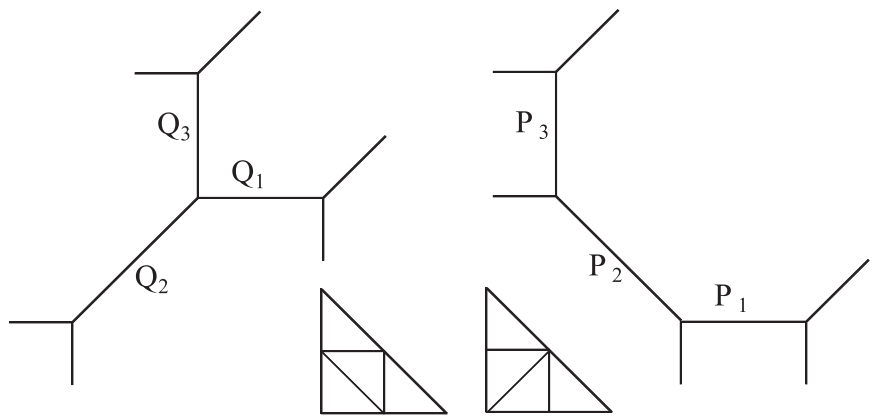} 
\begin{quote}
\caption{\emph{Two resolutions of the $\mathbb{C}^3/\mathbb{Z}_2\times\mathbb{Z}_2$ geometry, symmetric one (a.k.a. closed topological vertex, left) and asymmetric one (right), related by a flop of one $\mathbb{P}^1$. The K{\"a}hler parameters of both geometries
are related to each other \cite{cube} as $P_1=Q_1 Q_2, P_2=1/Q_2, P_3=Q_2 Q_3$.
}} \label{fig-CTV}
\end{quote}
\end{center}
\end{figure}

The corresponding Calabi-Yau manifold $\widetilde{Y}$ is a resolution of 
the orbifold $\mathbb{C}^3/
\mathbb{Z}_2 \times \mathbb{Z}_2$. 
There are two such resolutions, the symmetric one 
(also known as closed topological vertex) and the asymmetric one. 
Both of these resolutions consist of three
$\mathbb{P}^1$'s and are related to each other by a flop of one 
of the $\mathbb{P}^1$'s, see figure \ref{fig-CTV}. 
The appropriate geometry underlying our solution is the symmetric resolution.
Indeed, when $|Q|, |\mu|, |\epsilon| < 1$, 
the equation (\ref{symmetricresolution}) describes the mirror 
of the symmetrically resolved orbifold, with 
$Q, \mu, \epsilon^2$ being exponentials of flat coordinates of 
the K\"ahler moduli space, as we discuss in more detail below. 

We also note that on general grounds it is known that for Calabi-Yau manifolds of the form $uv+H(x,y)=0$, with $H(x,y)=0$ encoding a Riemann surface as in (\ref{symmetricresolution}), the special geometry relations reduce to
$$
T = \oint_a \lambda,\qquad  \frac{\partial F^{top}_0}{\partial T} = \oint_b \lambda,
$$
where $\lambda$ is a reduction of the holomorphic three-form along $u,v$ directions, $a$ and $b$ are dual one-cycles on a Riemann surface $H(x,y)=0$, and $F^{top}_0$ is the topological string free energy. The same relations hold for the free energy $F_0$ of matrix models, if $T$ is identified with the 't Hooft coupling \cite{Marino}. Therefore the fact that the spectral curve in the case we consider agrees with the mirror curve of $\widetilde{Y}$, ensures the agreement of derivatives of matrix model and topological string free energies with respect to $T$, up to an integration constant which is a function of $Q$ and $\mu$ (which are just parameters of the matrix potential). As the exact topological string partition function is a symmetric function of $Q$, $\mu$ and $\epsilon^2$, this implies that this integration constant must restore this symmetry, and the resulting matrix model free energy
$$
F_0 =  \textrm{Li}_3(Q)  +  \textrm{Li}_3(\mu) + \textrm{Li}_3(\epsilon^2) + \textrm{Li}_3(Q \mu \epsilon^2) - \textrm{Li}_3(Q\epsilon^2) - \textrm{Li}_3(\mu\epsilon^2)  - \textrm{Li}_3(Q\mu),
$$
has to agree with the topological string result $F^{top}_0$.

%It is also worth noting that the flop transition provides 
%a geometric way to explain why, upon varying chambers, 
%he value of $\mu=Q^{-1}e^{-\tau}\in [0,Q^{-1}]$ has to be limited by $Q^{-1}$.
%nder the flop the parameters $Q_1,Q_2,Q_3$ of the symmetric resolution 
%re related to $P_1,P_2,P_3$ of the asymmetric one as $P_1=Q_1 Q_2, P_2=1/Q_2, P_3=Q_2 Q_3$, 
%s discussed in \cite{cube}. To the leading order we can 
%onsider parametrization $Q_1\simeq \mu, Q_2\simeq Q,Q_3\simeq \epsilon^2$, 
%and then flop $\mathbb{P}^1$ related to $Q$, which results 
%n asymmetric resolution parametrized by $P_1\simeq e^{-\tau},P_2\simeq 1/Q,P_3\simeq Q\epsilon^2$. 
%From this viewpoint this is reasonable that $P_1$ cannot exceed 1, 
%ecause the local bundle of $P_1$ is of $\mathbb{C}^3/\mathbb{Z}_2$ 
%type which itself cannot undergo a flop. This is a geometric counterpart 
%of the fact that there are no negative chambers and negative values of $\tau$ are forbidden.

\begin{figure}[htb]
\begin{center}
\includegraphics[width=0.7\textwidth]{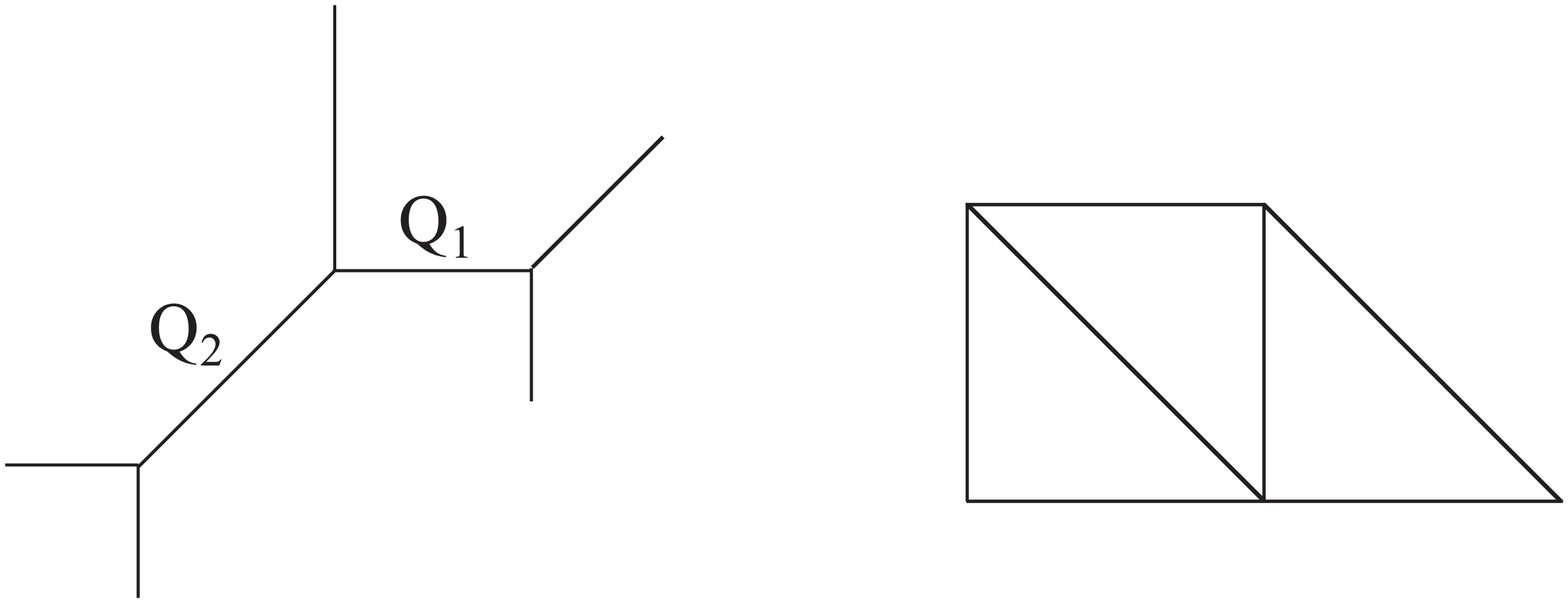} 
\begin{quote}
\caption{\emph{Toric diagram for the Suspended Pinch Point (SPP) geometry, and the corresponding dual diagram.  
This manifold contains two copies of $\mathcal{O}(-1,-1)\to\mathbb{P}^1$ geometry.  
}} \label{fig-SPP}
\end{quote}
\end{center}
\end{figure}

For the BPS counting problem, we are interested in the limit of 
$T \rightarrow \infty$, or equivalently $\epsilon \rightarrow 0$. 
With appropriate shifts of $x$ and $y$, the equation 
(\ref{symmetricresolution}) in this limit becomes
\be
  \mu\ e^{2y} + e^{x+y} + e^x + (1 + Q\mu)\ e^y + Q = 0.
\label{SPPcurve}
\ee
The manifold $Y$ corresponding to this curve is the SPP geometry,
with $Q$ and $\mu$ being exponentials of 
flat coordinates representing sizes of its two $\mathbb{P}^1$'s, 
which encode two copies of the initial $\mathcal{O}(-1,-1)\to\mathbb{P}^1$ geometry, see figure \ref{fig-SPP}. 
Not only does the spectral curve agree
 with the mirror curve of the SPP geometry in the 
limit of $g_s \rightarrow 0$, but in fact
the matrix integral reproduces the full topological string 
partition function at finite $g_s$.
Indeed, it is known that the SPP topological string partition function,
with K{\"a}hler parameters $Q$ and $\mu$, is equal to
\be
 Z^{\rm SPP}_{{\rm top}}(q, Q, \mu) 
= \prod_{k=1}^\infty \frac{(1-Qq^k)^k(1-\mu q^k)^k}
{(1-q^k)^{3k/2}(1 - \mu Q q^k)^k}.     \label{ZtopSPP}
\ee
On the other hand, from the explicit structure of the BPS generating
function and formulas (\ref{ZBPS-n-conifold}), (\ref{ZBPS-Zmatrix}) and (\ref{Cnconifold}),
we find that the value of the matrix integral, in the $N \rightarrow \infty$ limit, 
is related to the above topological string partition function as
\be   
  Z_{{\rm matrix}}(q, Q; n)  =  Z^{\rm SPP}_{{\rm top}}(q, Q, \mu = Q^{-1}q^n)
  \cdot \prod_{k=1}^\infty (1-q^k)^{k/2}.
 %& = & \prod_{k=1}^\infty (1-q^k)^k \cdot \prod_{j=1}^{\infty} \frac{(1-Qq^j)^j(1-\mu q^j)^j}{(1-\mu Q q^j)^j }
 \label{uptomacmahon}
\ee
In this way, the K\"ahler moduli $Q$ and the chamber number
$n$ for the BPS counting on the conifold are unified
into the two K\"ahler moduli of the SPP geometry. 
We note that there is an extra factor of the MacMahon function
in this relation. The appearance of the MacMahon factor, 
which is independent of $Q$, is a common and subtle issue
in relations between the topological string and other systems.

We note that the spectral curves 
(\ref{symmetricresolution}) and (\ref{SPPcurve})
arising from the matrix model automatically encode the relevant mirror map.
For example, in the parametrization of \eqref{SPPcurve}, 
$Q$ and $\mu$ are directly identified with the exponentials of the flat coordinates. We can verify this by explicit evaluation of period integrals, see \cite{IIKY}, section 3.3. 
The form (\ref{SPPcurve}) of the curve factorizes for $\mu=1$ and $Q=1$, which is consistent with degeneration of the topological
string partition function (\ref{ZtopSPP}) for these values. Also in the limit $x\to \pm\infty$, the solutions of the curve equation 
for $e^y$ reproduce appropriate locations of the asymptotic legs of the SPP toric diagram in figure \ref{fig-SPP}. 
The same parametrization naturally arises also in \cite{OY2} as a characteristic polynomial of the dimer model; see the example in section 4.2 and in particular (4.2.8) of \cite{Yamazaki}. 
All these arguments can be extended to the $\widetilde{Y}$ mirror curve (\ref{symmetricresolution}).
Note that the standard parametrization of this mirror curve, such as the
one in \cite{horivafa},  
would suggest the equation $e^{x+y} + e^x + e^y + \epsilon^2 e^{2x}  + \mu e^{2y} + Q= 0$. This is however valid for large values of 
K{\"a}hler parameters, and consistent with  (\ref{symmetricresolution}),
as in this regime the quadratic terms in $Q,\mu$ and $\epsilon^2$ are negligible.

Because of the form of the spectral curve at finite 't Hooft coupling
(\ref{symmetricresolution}), it is natural to conjecture
that the partition function of our conifold matrix model
for finite 't Hooft coupling is equal to the topological
string partition function of the resolution of
$\mathbb{C}^3/\mathbb{Z}_2 \times \mathbb{Z}_2$ \cite{cube}, 
modulo a MacMahon factor
\be
 Z^{\rm total}_{{\rm matrix}}(q, Q, \mu,\epsilon^2) 
= \prod_{k=1}^\infty (1-q^k)^k \cdot \prod_{k=1}^\infty \frac{(1-Qq^k)^k(1-\mu q^k)^k (1-\epsilon^2 q^k)^k(1-Q\mu \epsilon^2 q^k)^k}
{(1 -  Q\mu q^k)^k (1 - \mu \epsilon^2 q^k)^k (1 - Q\epsilon^2 q^k)^k}.    \label{Ztotal}
\ee
We chose the MacMahon factor in such a way that it reduces to
our result (\ref{uptomacmahon}) in the infinite 't Hooft coupling
limit $\epsilon\to 0$.
As another evidence for the conjecture, we point out that, 
in the limit $Q,\mu\to 0$, our model reduces to the 
Chern-Simons matrix model (discussed in the next section) 
and the above partition function correctly reduces to the
appropriate Chern-Simons partition function. 
It would be interesting to test this conjecture, for example 
by applying matrix model recursion relations of \cite{EO}. 

As discussed in  \cite{cube}, the right-hand side
of (\ref{Ztotal}) is precisely (including the correct power of
MacMahon function) the generating function of plane partitions in a finite $K\times L\times M$ cube,
and up to one power of MacMahon reproduces the closed topological vertex partition function with 
K{\"a}hler parameters identified as $Q_1=g_s K, Q_2=g_s L, Q_3=g_s M$ (this generalizes $\bC^3$ model 
of plane partitions with one wall discussed in section \ref{ssec-C3matrix}). 
In the present case we have one analogous identification of the 't Hooft parameter $T=g_s N$. 
Our 't Hooft parameter has also a nice combinatorial interpretation: finite $N$ corresponds to matrices with $N$ eigenvalues, 
which in the construction of our matrix models arise from truncation of products in (\ref{I-matrix}) 
to $N$ operators $\Gamma'$. This translates
to truncation of Young diagrams, which arise from slicing of the crystal model pyramid, to at most $N$ rows, 
which is equivalent to considering a wall at location $N$. Therefore our present model involves one wall 
associated to finite 't Hooft coupling, the second parameter $\mu$ which involves finite $n$ (which also measures
a size of the crystal), and the third parameter $Q$ which appears in the matrix model potential in the same way
as $\mu$, however does not have a clear crystal interpretation. The cube model of \cite{cube} involves three symmetric walls
and has the same generating function (\ref{Ztotal}). It would be interesting to understand 
the relations between these two models in more detail.

As the final remark, we note that there are three limits in which our full matrix model reproduces both
the mirror curve, as well as the topological string partition function of the conifold. 
The first such limit $\mu,Q\to 0$ brings us to the Chern-Simons matrix model and will be discussed
in the next section. The second limit $\mu,\epsilon\to 0$ is just the commutative limit of the model 
with matrices of infinite size. In both these limits it is not surprising that the size of the conifold
is identified respectively with 't Hooft coupling $e^{-T}$ or the original K{\"a}hler parameter $Q$. 
However in the third limit $Q,\epsilon\to 0$ we obtain the conifold of the size $\mu=Q^{-1}e^{-\tau}$, which
in fact means the $Q$ vanishes however the chamber parameter $\tau\to \infty$. It also corresponds to 
the commutative limit, and shows that for vanishing $Q$ the role of the conifold K{\"a}hler parameter
is attained by $\mu$. This is in agreement with the picturesque identification of the conifold size
with the length of the top row of the pyramid in the crystal melting model, and puts this
identification on the firmer footing.

\subsection{Relation to the Chern-Simons Matrix Model}    \label{ssec-CSmm}

We now discuss the commutative chamber $n\to\infty$ of the conifold model presented above. We show that it leads to the 
matrix model which is equivalent to the Chern-Simons matrix model, and these two models can be unified in a geometric way. By the Chern-Simons matrix model \cite{mm-lens,CSmatrix,Marino} we understand the unitary matrix model with the Gaussian potential, 
as in (\ref{Gaussian-C3}), and finite 't Hooft coupling $T=Ng_s$.
Including the shift (\ref{measure-T}) arising from the measure, we write its potential as
%\footnote{The negative sign of the quadratic term is a consequence of our conventions, 
%which relate to those in \cite{CSmatrix,Marino} by the change of the sign of $g_s$.}  
\be
V_{CS}= T \log U- \frac{1}{2}(\log U)^2, \qquad \qquad \partial_U V_{CS} = \frac{T - \log U}{U},   \label{oldCS} %
\ee
%This potential is written in terms of variable $z=e^u$, for which the unitary measure 
%can be turned into the standard Vandermonde determinant at the expense 
%of introducing an additional  linear term $(T\log z)$ to the potential, 
%with 't Hooft coupling $T=Ng_s$. 

\begin{figure}[htb]
\begin{center}
\includegraphics[width=0.4\textwidth]{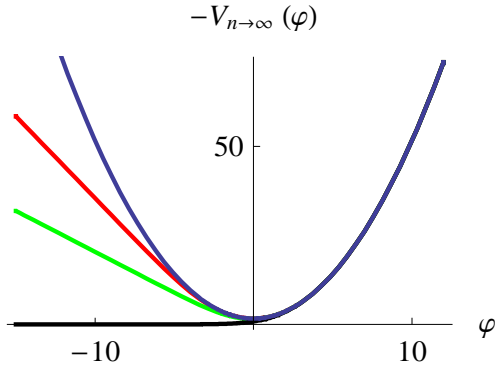} 
\begin{quote}
\caption{\emph{Matrix potential (without (\ref{measure-T}) shift), $-V_{n\to\infty}(\varphi) = \frac{\pi^2}{6}+\frac{\varphi^2}{2}+{\rm Li}_2(-Qe^{-\varphi})$ in terms of a variable $u=e^{\varphi}$. The blue plot represents the Gaussian potential with $Q=0$. Increasing $Q$ flattens the potential (red and green). At the conifold singularity, corresponding to $Q=1$, the potential becomes flat (black plot).
}} \label{fig-Vcommute}
\end{quote}
\end{center}
\end{figure}

Our present model is also unitary and in the commutative chamber the derivative of its potential (\ref{Vconifold}) reduces to
\be
\partial_U V_{n \to\infty} = \frac{T - \log(U+Q)}{U}.     \label{Gaussian-Q}
\ee
We recall that our matrix model arises from rewriting the BPS generating function, which in the $n$-th chamber takes form (\ref{ZBPS-n-conifold}).
In the commutative chamber $n\to\infty$ the term $M(Q^{-1})=\prod_k (1-Q^{-1}q^k)^{-k}$ is removed from that expression. On the other hand, in this limit the prefactor (\ref{Cnconifold}) reduces to a single MacMahon function $M(1)=\prod_k (1-q^k)^{-k}$. Therefore in the commutative chamber we find
$$ 
\frac{M(1)}{M(Q)} = 
\int \prod_i du_i \, \prod_{j<k}(u_j-u_k)^2  \, \prod_k e^{-\frac{1}{g_s} V_{n\to\infty}(u_i)}.
$$
The ratio on the left hand side is precisely the partition function of the Chern-Simons theory on $S^3$, which is also reproduced by the Chern-Simons matrix model (\ref{oldCS}).  The spectral curve of that model has genus zero and is identified with $\mathbb{P}^1$ which arises from the geometric transition of the $S^3$. The size of this $\mathbb{P}^1$ is given by the (finite) 't Hooft coupling $T$. Now we find the model whose partition function is given by the same Chern-Simons partition function and its spectral curve has also genus zero, however our association of parameters is different. Instead of finite 't Hooft coupling parameterizing the size of $\mathbb{P}^1$, in our model 't Hooft coupling is infinite, while the size of $\mathbb{P}^1$ is encoded in a fixed parameter $Q$ deforming the unitary Gaussian potential as in (\ref{Gaussian-Q}). As an immediate check we notice that for $Q=0$ our potential (\ref{Gaussian-Q}) indeed reduces to (\ref{oldCS}), and for infinite $T$ it reproduces Gaussian result for plane partitions (\ref{Gaussian-C3}). The dependence of the potential $V_{n\to\infty}$ on the parameter $Q$ is shown in figure \ref{fig-Vcommute}. 
As we approach the conifold singularity at $Q=1$, it is interesting to observe the flattening of the matrix potential. It is known that the conifold singularity has to do with the flattening of the Coulomb branch moduli space \cite{Witten:1993yc, Ooguri:2002gx}. This indicates a connection between the matrix variable and the Coulomb branch variables. 

With both finite 't Hooft coupling $T$ and finite $Q$, we find a unifying geometric viewpoint, again in terms of the SPP geometry, however now with K{\"a}hler parameters $Q$ and $e^{-T}$.
%This Gaussian potential, together with the finite 't Hooft coupling, 
%is again the content of the old Chern-Simons matrix model \cite{mm-lens,lens-matrix,Marino}.
In this topological string limit the equations (\ref{ab-eq1}) and (\ref{ab-eq2}) take form
\bea
\sqrt{a_+ +Q} + \sqrt{a_- +Q} & = & 2e^{-T/2}, \\
\sqrt{a_-(a_+ +Q)} + \sqrt{a_+ (a_-+Q)} & = & (\sqrt{a}+\sqrt{b})e^{T/2}, 
\eea
and their solution is given by
\be
a_{\pm} = -1 + (2-Q)\epsilon^2 \pm 2i\epsilon \sqrt{(1-Q)(1-\epsilon^2)},   \label{aTopString} 
%b_{-} = -1 + (2-Q)\epsilon^2 - 2i\epsilon \sqrt{(1-Q)(1-\epsilon^2)},   \label{bTopString} 
\ee
%The resolvent is of the form (\ref{omega-coni}) with
%$$
%\gamma_1^{top.string} = \frac{b\sqrt{a+Q} - a\sqrt{b+Q}}{\sqrt{a+Q} - \sqrt{b+Q}} 
%= -1-Q\epsilon^2,\qquad\qquad \gamma_2^{top.string}=1.
%$$
%Equivalently, for $\mu=0$ the resolvent given in (\ref{omega-coni-muSmall}) reduces to
which leads to the following form of the resolvent 
\be
\omega(u)_{\mu=0} = \frac{1}{uT} \log \Big(\frac{u + 1+Q e^{-T}  -  \sqrt{(u+1+Qe^{-T})^2-4(u+Q)\epsilon^2}}{2 e^{-T} (u+Q)} \Big).   \label{omega-coni-mu0}    
\ee
The spectral curve which arises from this resolvent is again mirror curve of the SPP geometry and
%(and which of course coincides with $\mu=0$ case of (\ref{symmetricresolution})) 
reads
\be
x + u + xu + x^2 \frac{\epsilon^2}{1+Q\epsilon^2} 
+ \frac{Q}{1+Q\epsilon^2} = 0. \label{curveSPPRedef}
\ee
It is clear that $\epsilon\to 0$ limit leads to the conifold geometry with the conifold of size $Q$. Finally, for $Q=0$
the resolvent (\ref{omega-coni-mu0})
$$
\omega(u)_{\mu=Q=0} = \frac{1}{uT} \log \Big(\frac{u + 1 - \sqrt{(u+1)^2- 4u e^{-T}}}{2u e^{-T}} \Big)   
$$
agrees\footnote{Instead of introducing $T\log U$ term to the potential (\ref{oldCS}) to get the standard Vandermonde determinant, the solution in \cite{Marino} involves completing the square, which leads to a redefinition $u_{\rm here}=p_{\textrm{\cite{Marino}}}e^T$. Due to a different sign of $g_s$ we also need to identify 't Hooft couplings as $T_{\rm here}=-t_{\textrm{\cite{Marino}}}$. Taking this into account, our cut endpoints (\ref{aTopString}) with $Q=0$ also agree with those in \cite{Marino}.} with the resolvent of the Chern-Simons matrix model found in \cite{CSmatrix,Marino}, and the spectral curve reproduces the conifold mirror curve of the size given by the 't Hooft coupling
$$
x + u + xu + x^2 e^{-T}  = 0. 
$$

%*******************************************************************

\subsection{$\mathbb{C}^3/\mathbb{Z}_2$}

\begin{figure}[htb]
\begin{center}
\includegraphics[width=0.7\textwidth]{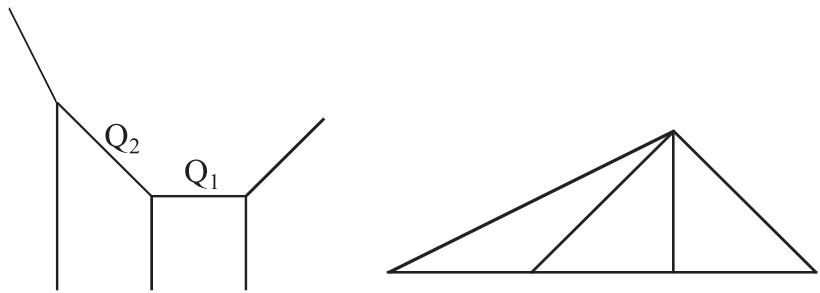} 
\begin{quote}
\caption{\emph{Toric diagram for the resolution of $\mathbb{C}^3/\mathbb{Z}_3$ singularity, and the corresponding dual diagram.  
This geometry contains two copies of $\mathbb{C}^3/\mathbb{Z}_2$ resolution.
}} \label{fig-C3Z3}
\end{quote}
\end{center}
\end{figure}

A similar analysis as for the conifold can be performed for $\mathbb{C}^3/\mathbb{Z}_2$
geometry, for arbitrary chamber. Even though we do not repeat a matrix model derivation of the spectral curve for this case,
we note that the relation to topological string theory is also immediate, and the relevant 
geometry $Y$ for this case is the resolution of $\mathbb{C}^3/\mathbb{Z}_3$ singularity shown in figure \ref{fig-C3Z3}.
This geometry contains two $\mathbb{P}^1$'s of $\mathcal{O}(0,-2)$ type, and denoting its K{\"a}hler parameters
by $Q$ and $\mu$, its topological string partition function reads
\be
 Z^{\mathbb{C}^3/\mathbb{Z}_3}_{{\rm top}}(q, Q, \mu) 
= \prod_{k=1}^\infty (1-q^k)^{-3k/2} (1-Qq^k)^{-k}(1-\mu q^k)^{-k}
(1 - \mu Q q^k)^{-k}.
\ee
Therefore, from (\ref{matrixorbifold}) and (\ref{CnZ2}) we find in this case
\be   
  Z_{{\rm matrix}}(q, Q; n)  =  Z^{\mathbb{C}^3/\mathbb{Z}_3}_{{\rm top}}(q, Q, \mu = Q^{-1}q^n)
  \cdot \prod_{k=1}^\infty (1-q^k)^{k/2} .
 \label{uptomacmahonC3Z2}
\ee
This shows that the matrix model partition function (\ref{matrixorbifold}) 
in the $n$-th chamber is equal to the topological string partition 
function for
$\mathbb{C}^3/\mathbb{Z}_3$ with its two K\"ahler moduli given by
$Q$ and $\mu = Q^{-1} q^n$, up to the MacMahon function as 
in (\ref{uptomacmahon}). This unified geometry $Y$ contains two copies of the initial
$\mathbb{C}^3/\mathbb{Z}_2$ resolution. It would be interesting to check what 
geometry $\widetilde{Y}$ would arise for finite 't Hooft coupling, and 
whether it is consistent with the total matrix model partition function.

\subsection{General Toric Calabi-Yau Manifold}

Matrix model for a general toric manifold and in general chamber can be constructed in a similar manner, following fermionic approach
of \cite{2009pyramid,NagaoVO}, and then analyzed along the lines above. We do not present a construction in a general chamber which is
technically much more involved, however we found explicit expressions for matrix models for a general manifold $X$ in the non-commutative chamber. These models are presented in section \ref{subsec.generalmatrix}. Nonetheless, we postulate that for arbitrary chamber, in the infinite 't Hooft limit, we should also find a toric manifold $Y$ which contains two copies of $X$. For finite 't Hooft limit matrix model spectral curve would encode yet more general manifold $\widetilde{Y}$.

%%%%%%%%%%%%%%%%%%%%%%%%%%%%%%%%%%%%%%%%%%%%%%%%%%%%%%%%%%%%%%%%%%%%%%%%%
\section{Derivations of the Matrix Models}

In this section we give two derivations of our matrix models. 
One derivation (section \ref{subsec.derivation1}) uses the free fermion formalism, while the other (section \ref{subsec.derivation2}) uses a set of non-intersecting paths.\footnote{We have been informed by Mina Aganagic
that there is yet another derivation of
the matrix models \cite{mina}.} 

%The relations between the two derivations will be clarified in section \ref{subsec.relation}.

Both derivations are based on the following observation.
Let us begin with the crystal melting model of \cite{OY1}. Given a configuration of a crystal, we can slice the crystal by a sequence of parallel planes. On each slice, we have a Young diagram. The Young diagrams evolve according to the interlacing conditions, which is equivalent to the melting rules of \cite{OY1}. 

For $\bC^3$, we have \cite{ORV}
\beq
\ldots \prec \lambda(-2) \prec \lambda(-1) \prec \lambda(0) \succ \lambda(1) \succ \lambda(2) \succ\ldots
\label{eq.lambdaC3interlacing}
\eeq
where we write $\lambda\succ\mu$ (equivalently $\mu\prec \lambda$) 
for two partitions $\lambda=(\lambda_i)$ and $\mu=(\mu_i)$, if
\beq
\lambda_i=\mu_i+1 \quad \mathrm{or} \quad \lambda_i=\mu_i \quad \textrm{for each}~~ i.          \label{def-succ}
\eeq
For conifold in the non-commutative chamber \cite{YoungBryan},
\beq
\ldots \prec \lambda(-2) \pl \lambda(-1) \prec \lambda(0) \pg \lambda(1) \succ \lambda(2) \pg \ldots
\label{eq.lambdaconifoldinterlacing}
\eeq
where we write $\lambda\pg\mu$ for $\lambda=(\lambda_i), \mu=(\mu_i)$ if $\lambda^t \succ\mu^t$, $i.e.$
\beq
\ldots \geq \lambda_i\geq \mu_i \geq \lambda_{i-1}\geq \mu_{i-1} \ldots    \label{def-pg}
\eeq

\begin{figure}[htb]
\begin{center}
\includegraphics[width=0.9\textwidth]{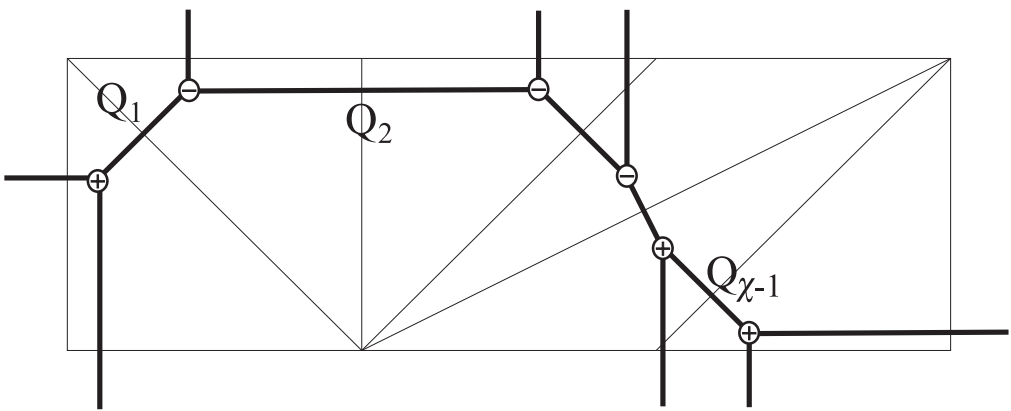} 
\begin{quote}
\caption{\emph{Toric diagram for Calabi-Yau manifold without compact 
4-cycles arises from a triangulation of a strip. There are $N$ independent $\mathbb{P}^1$'s with K{\"a}hler parameters $Q_i=e^{-t_i}$, and $\chi$ vertices to which we associate $\oplus$ and $\ominus$ signs $S_i$. Intervals which connect vertices with opposite signs represent $\mathcal{O}(-1,-1)\to \mathbb{P}^1$ local neighborhoods. Intervals which connect vertices with the same signs represent $\mathcal{O}(-2,0)\to \mathbb{P}^1$ local neighborhoods. 
%%The first vertex on the left is chosen to be $\oplus$.
}} \label{fig-strip}
\end{quote}
\end{center}
\end{figure}

We can also discuss more general toric Calabi-Yau 3-folds $X$ without compact 4-cycles (see figure \ref{fig-strip}).
The $(p,q)$-web for $X$ has $\chi$ vertices, where $\chi$ is the Euler characteristics of $X$.
To each vertex we associate a sign $S_i=\pm 1$ so that, 
% we want to keep t_i for Kahler moduli
\beq
S_i S_{i+1}=s_i,
\eeq
where the sign factor $s_i = \pm 1$ is defined in section 2.4. 
This means that if the local neighborhood of $i$-th $\mathbb{P}^1$ 
represented by an interval between vertices $i$ and $i+1$ 
is $\mathcal{O}(-2,0)$, then $S_{i+1}=S_i$; if this neighborhood 
is of $\mathcal{O}(-1,-1)$ type, then $S_{i+1}=-S_i$.
There is a binary choice of overall signs $S_i$: the type of the first 
vertex could be chosen as either $S_1=+1$ or $S_1=-1$. This choice corresponds to the exchange of rows and columns of Young diagrams. Each choice
gives rise to a matrix model potential, and they are related to
each other by analytic continuation.

Given such $S_i$, the interlacing conditions in the noncommutative chamber are given by
\beq
\ldots \overset{S_{-2}}{\prec} \lambda(-2) \overset{S_{-1}}{\prec} \lambda(-1) \overset{S_0}{\prec} \lambda(0) \overset{S_1}{\succ} \lambda(1) \overset{S_2}{\succ} \lambda(2) \overset{S_3}{\succ} \ldots
\eeq
where $ \overset{-}{\succ} = \succ$
and we extended the definition of 
$S_i~ (i=1,\ldots, \chi)$ to $S_i ~(i\in \bZ )$ by periodic 
identification: $S_{i+\chi}=S_i$. More general expression, applicable to any chamber, is given in \cite{NagaoVO,NY}.

\subsection{Derivation (I): Free Fermions}\label{subsec.derivation1}

\subsubsection{Wall Crossing and Free Fermions}\label{subsec.fermionreview}

The first derivation is based on the free fermion formalism developed in \cite{2009pyramid,NagaoVO} (see also \cite{NY}), which we now review briefly.

The basic idea is as follows. We have seen that states in the crystal melting model are represented by Young diagrams. Since Young diagrams are represented by states of free fermion systems and evolutions of Young diagrams by vertex operators,  the partition function is written as a correlator of fermions bilinears \cite{ORV,YoungBryan}. 

We give the resulting expression in the notation of \cite{2009pyramid}.
For any toric geometry without compact 4-cycles, the generating function of DT invariants in the non-commutative chamber can be written as
\be
Z_{\rm BPS}(q,Q;n=0) = \langle \Omega_+ | \Omega_- \rangle,     \label{ZOmega}
\ee
where $|\Omega_{\pm}\rangle$ are fermionic states which will be defined below.
Moreover, we can introduce wall crossing operators $\overline{W}_p$ 
to write expression in other chambers,\footnote{In this example, inserting wall crossing operators is equivalent to commuting vertex operators, which is proposed in \cite{NagaoVO,NY}.} where $n_p=m$ for the $p$-th $\bP^1$ and all other $n$'s set equal to zero:
\be
Z_{\rm BPS}\left(q,Q;n_p=m, \textrm{ all other $n=0$}\right) = \langle \Omega_+ | (\overline{W}_p(1))^m |\Omega_- \rangle.    \label{ZOmegaW}
\ee
In the remainder of this subsection we give explicit expressions for the states $|\Omega_{\pm}\rangle$ and wall-crossing operators $\overline{W}_p$.

We first define a vertex operator 
$\G_{\pm}^{S_i}(x)$ at each vertex as
$$
\G_{\pm}^{S_i=+1}(x)=\G_{\pm}(x),\qquad \qquad \G_{\pm}^{S_i=-1}(x)=\G_{\pm}'(x),
$$
where $\G_{\pm}(x)$ and $\G_{\pm}'(x)$ are defined in appendix \ref{app-fermion}. These operators represent the evolution of Young diagrams $\lambda(t)$, and $\Gamma_+, \Gamma_-, \Gamma'_+$ and $\Gamma'_-$ are nothing but the evolution rules $\prec, \succ, \pl$ and $\pg$, respectively \cite{YoungBryan}. 

Next, we consider a product of $\chi$ such operators $\G_{\pm}^{S_i}(x)$ interlaced with $\chi$ operators $\widehat{Q}_i$ representing colors $q_i$, for $i=0,1,\ldots,\chi-1$. Operators $\widehat{Q}_1,\ldots,\widehat{Q}_{\chi-1}$ are associated to $\mathbb{P}^1$ in the toric diagram (also defined in appendix \ref{app-fermion}), and there is an additional $\widehat{Q}_0$.
We then introduce
\be
\overline{A}_{\pm}(x) = \G_{\pm}^{S_1} (x) \widehat{Q}_1 \G_{\pm}^{S_2} (x) \widehat{Q}_2 \cdots \G_{\pm}^{S_{\chi-1}} (x) \widehat{Q}_{\chi-1} \G_{\pm}^{S_{\chi}} (x) \widehat{Q}_0.                  \label{Apm}
\ee
Commuting all $\widehat{Q}_i$'s to the left or right we also introduce 
\bea
A_+(x) & = & (\widehat{Q}_0 \widehat{Q}_1\cdots \widehat{Q}_{\chi-1})^{-1} \, \overline{A}_{+}(x) = \G_{+}^{S_1} \big(xq\big)  \G_{+}^{S_2} \big(\frac{xq}{q_1}\big)  \cdots \G_{+}^{S_{\chi}} \big(\frac{xq}{q_1q_2\cdots q_{\chi-1}}\big), \label{Aplus} \\
A_-(x) & = & \overline{A}_{-}(x) \, (\widehat{Q}_0 \widehat{Q}_1\cdots \widehat{Q}_{\chi-1})^{-1} = \G_{-}^{S_1} (x)  \G_{-}^{S_2} (xq_1)  \cdots \G_{-}^{S_{\chi}} (x q_1q_2 q_{\chi-1}). \label{Aminus}
\eea
The states $|\Omega_{\pm}\rangle$ are now defined as
\bea
\langle \Omega_+| & = & \langle 0 | \ldots \overline{A}_+(1) \overline{A}_+(1) \overline{A}_+(1) = \langle 0 | \ldots A_+(q^2) A_+(q) A_+(1),  \label{Omega-plus}  \\
| \Omega_- \rangle & = & \overline{A}_-(1) \overline{A}_-(1) \overline{A}_-(1) \ldots |0\rangle = A_-(1) A_-(q) A_-(q^2) \ldots |0\rangle .  \label{Omega-minus}
\eea
It was shown in \cite{2009pyramid} that we have the relation \eqref{ZOmega}
under the following identification between $q_i$ parameters which enter a definition of $|\Omega_{\pm}\rangle$ and topological string parameters $Q_i=e^{-T_i}$ and $q=e^{-g_s}$:
\be
q_i = (S_i S_{i+1}) Q_i,\qquad \qquad q=  q_0 q_1\cdots q_{\chi-1}.   \label{qQ}
\ee
In addition the wall-crossing operators are defined by
\begin{align}
& \overline{W}_p(x) =  \Big(\G_{-}^{t_1} (x) \widehat{Q}_1 \G_{-}^{t_2} (x) \widehat{Q}_2 \cdots \G_{-}^{t_p} (x) \widehat{Q}_p\Big)\Big( \G_{+}^{t_{p+1}} (x) \widehat{Q}_{p+1} \cdots \G_{+}^{t_{\chi-1}} (x) \widehat{Q}_{\chi-1} \G_{+}^{t_{\chi}} (x) \widehat{Q}_0 \Big)  \label{Wp} 
%\\
%& \overline{W}_p'(x)  =  \Big(\G_{+}^{t_1} (x) \widehat{Q}_1 \G_{+}^{t_2} (x) \widehat{Q}_2 \cdots \G_{+}^{t_p} (x) \widehat{Q}_p\Big)\Big( \G_{-}^{t_{p+1}} (x) \widehat{Q}_{p+1} \cdots \G_{-}^{t_{\chi-1}} (x) \widehat{Q}_{\chi-1} \G_{-}^{t_{\chi}} (x) \widehat{Q}_0 \Big) ,   \label{Wpprim}  
\end{align}
and the relation (\ref{ZOmegaW}) holds under the change of variables
\be
Q_p = (S_p S_{p+1})q_p q^{m},\qquad  Q_i = (S_i S_{i+1})q_i \quad \textrm{for}\quad i\neq p,\qquad  q=  q_0 q_1\cdots q_{\chi-1}. \label{ZRposBpos-qQ-Results}
\ee

%*******************************************************************
%*******************************************************************

\subsubsection{Matrix Models from Free Fermions}    \label{ssec-MatrixFermions}

Once the BPS partition function is written in the fermionic formalism, it can be turned into a matrix model upon inserting appropriately chosen identity operator in the correlator (\ref{ZOmegaW}):  
\be
Z_{\rm BPS}\left(q,Q;n_p=m, \textrm{ all other $n=0$}\right)  = \langle \Omega_+ |~ \mathbb{I} ~ (\overline{W}_p(1))^m |\Omega_- \rangle.       \label{ZOmega-I}
\ee
The identity operator $\mathbb{I}$ is represented by the complete set of states $|R\rangle\langle R|$ (representing two-dimensional partitions). Using orthogonality relations of $U(\infty)$ characters $\chi_R$, and the fact that these characters are given in terms of Schur functions $\chi_R=s_R(\vec{u})$ 
for $\vec{u}=(u_1,u_2,u_3,\ldots)$, we can write
\bea
\mathbb{I} & = & \sum_R |R\rangle\langle R|  = \sum_{P,R} \delta_{P^t R^t} |P\rangle\langle R| = \nonumber \\
& = & \int dU \sum_{P,R} s_{P^t}(\vec{u}) \overline{s_{R^t} (\vec{u})} |P\rangle\langle R| = \nonumber \\
& = & \int dU \Big(\prod_i \G_-'(u_i)|0\rangle  \Big)  
\Big(\langle0 | \prod_i \G_+'(u^{-1}_i) \Big),   \label{I-matrix}
\eea
where $dU$ denotes the unitary measure for $U(\infty)$, 
which can be written in terms of eigenvalues $u_i = e^{i\phi_i}$
of $U$ as,
$$
dU=\prod_k d\phi_k \,\prod_{i<j}\left(e^{i\phi_i} - e^{i\phi_j}\right)
\left(e^{-i\phi_i} - e^{-i\phi_j}\right).
$$

Having inserted the identity operator in this form into (\ref{ZOmega-I}) 
we can commute away $\G_{\pm}'$ operators and get rid of operator expressions. 
This leads to a matrix model with the unitary measure $dU$. 
In case of the non-commutative chamber all factors arising 
from commuting these $\G_{\pm}'$ operators depend on $u_i$ and contribute 
just to the matrix model potentials. In other chambers  additional 
factors arise which do not depend on $u_i$ and therefore contribute 
to some overall factor $C_n$ (in a chamber labeled by $n$). 

Thus in general we write the DT generating function as 
a matrix model, up to the factor $C_n$. 
In the non-commutative chamber, the integrand
can be expressed in terms of the theta-product,
$$
 \Theta(U|q) = \prod_{k=0}^{\infty} (1 + Uq^k)(1+U^{-1}q^{k+1}),
$$
and in other chambers of certain modification thereof.

We emphasize here that this fermionic method of constructing matrix models 
applies to any chamber for any toric Calabi-Yau 3-fold without 
compact 4-cycles. This includes, for example, chambers where the 
BPS partition function becomes a finite product \cite{2009pyramid}.

\bigskip

One may ask if our construction of the matrix model is unique. 
One potential source of ambiguity
is the location of the operator $\mathbb{I}$.
In \eqref{ZOmega-I} we inserted the operator $\mathbb{I}$ on 
the left side of $(\overline{W}_1(1))^m$. When inserted on the right, 
we find a seemingly different matrix model potential,
for example,
\be
 \int dU\ {\rm det}\left( \prod_{k=0}^\infty \frac{(1+U q^k)\ 
 (1+U^{-1} q^{k+n+1})}{(1+QU q^{k-n}) (1+Q^{-1}U^{-1}q^{k+n+1})}\right),
\ee
in the conifold with the same prefactor (\ref{Cnconifold}). 
This integral can be turned into
\be
 Z_{{\rm matrix}}(q,Q;n) = 
\int dU\ {\rm det}\left(\frac{\Theta(QU|q)}{\Theta(U|q)}
  \prod_{k=1}^{n}\frac{1}{1+Q^{-1}U^{-1}q^k}\right),
\label{conifoldmatrix}
\ee
by the simple change of integration variable, 
$U \rightarrow q^{n+1}Q^{-1} U^{-1}$. 
The resulting integral is similar to our original integral (\ref{conifoldmatrixone}),
but the numerator and the denominator are exchanged. 
Then the matrix model (\ref{conifoldmatrix}) can be derived by taking advantage of
the freedom of changing overall signs $S_i \rightarrow - S_i$'s, which was mentioned in section 
\ref{subsec.fermionreview}. 
The two matrix models can then be related
by analytic continuation in $q$, as described
in the footnote 5 of \cite{Ooguriknot}.
Another possible ambiguity involves renaming $U\to U^{-1}$ in (\ref{I-matrix}),
which does not affect the form of the measure, 
however may affect the form of the potential.

%%%%%%%%%%%%%%%%%%%%%%%%%%%%%%%%%%%%%%%%%%%%%%%%%%%%%%%%%%%%%%%%%%%%%%%%%%%%%
\subsection{Derivation (II): Non-intersecting Paths}
\label{subsec.derivation2}

In this section we give yet another derivation of our unitary matrix models, based on the non-intersecting paths. The fundamental observation here is that states in the crystal melting model, $i.e.$, a sequence of Young diagrams satisfying interlacing conditions, can be equivalently expressed as a set of non-intersecting paths on an oriented graph.\footnote{The matrix model for $\bC^3$ is constructed recently by \cite{EynardTASEP}. We will generalize those arguments to conifold later. Also, even for $\bC^3$ the explicit expression of the potential for the 1-matrix model \eqref{eq.renomV} in our paper seems to be new.} 
Using the Linstr\"om-Gessel-Viennot (LGV) formula described in Appendix C, 
we can express the result as a determinant, which is an integral over a matrix whose eigenvalues are the height of the non-intersecting paths. 
%Part of the evolution rules of Young diagrams are implemented by delta functions, which in turn can be traded for Lagrange multipliers. 
This gives a multi-matrix model. We finally simplify the matrix model into a 1-matrix model.

\subsubsection{$\bC^3$}

Let us begin with the simplest example, $\bC^3$. Let us define
\beq
h_k(t)=\lambda_{N-k+1}(t)+k-1,
\label{eq.hdef}
\eeq
for $k=1,\ldots, N$. Since $\lambda(t)$ is a partition, we have
\beq
h_k(t)< h_{k+1}(t),
\label{eq.hnonintersecting}
\eeq
for all $t$. We also have the boundary condition,
\beq
h_k(t)=k-1 ~\mathrm{when} ~|t| ~\textrm{large} .
\label{eq.hboundary}
\eeq
Moreover, \eqref{eq.lambdaC3interlacing} means we have, for each step $t$,
\begin{equation*}
h_k(t+1)-h_k(t)= 0 ~{\rm or}\, -1,
\end{equation*}
for $t \geq 0$ and 
\begin{equation*}
h_k(t+1)-h_k(t)= 0 ~{\rm or}~ 1,
\end{equation*}
for $t < 0$. For later purpose, it is convenient to
introduce the minus of the sign function
$\sigma(t') ~(t'\in \bZ+\frac{1}{2})$, 
\beq
\sigma(t')=
\begin{cases}
-1 & (t'> 0). \\
+1 & (t'<0).
\end{cases}
\eeq
so that
\be
h_k(t+1)-h_k(t)= 0 ~{\rm or}~ \sigma(t+\frac{1}{2}),
\label{eq.hevolve}
\ee
Summarizing, we see that Young diagrams $\{\lambda(t) \}$ are equivalently expressed by a set of coordinates $h_k(t)$, which satisfy the conditions above.

\begin{figure}[htbp]
\centering{\includegraphics[scale=0.35]{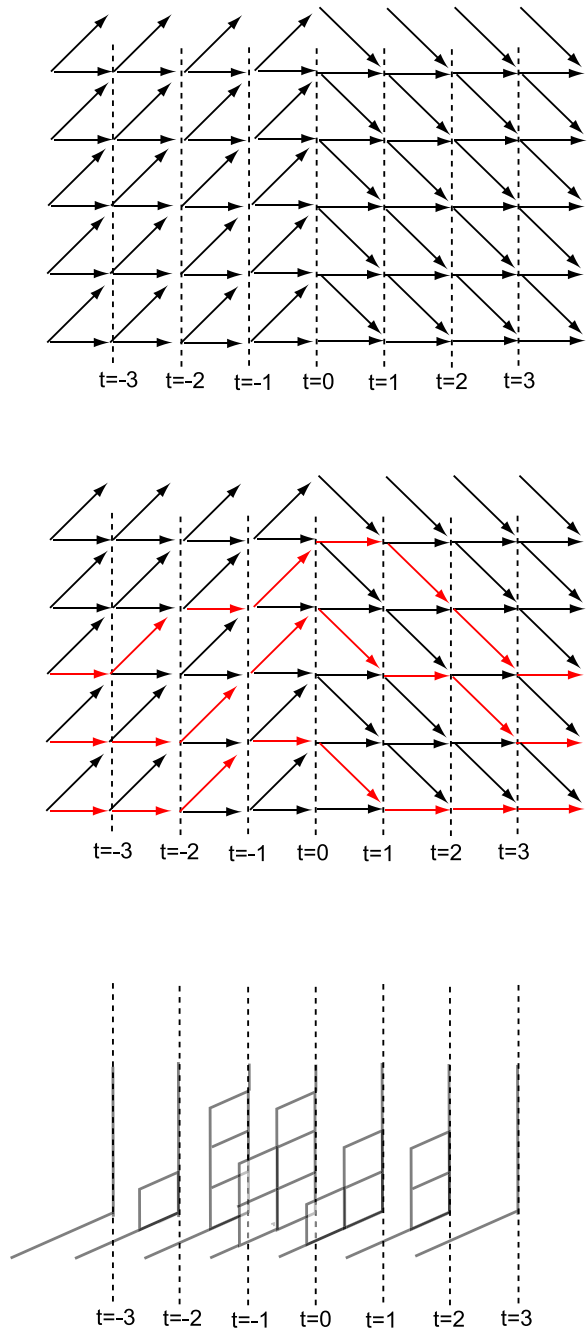}}
\begin{quote}
\caption{\emph{Top: an oriented graph for $\bC^3$. Middle:
an example of 3 non-intersecting paths shown in red. 
The location of the $k$-th path at time $t$ gives $h_k(t)$. Bottom: The corresponding evolution of Young diagrams.}}
\label{fig.C3orientedgraph}
\end{quote}
\end{figure}

We can represent these coordinates as a set of non-intersecting paths on an oriented graph shown in figure \ref{fig.C3orientedgraph}.\footnote{This oriented graph arises from the lozenge tiling of the plane, which is another way of representing crystal for $\bC^3$. The paper \cite{EynardTASEP} uses this tiling to construct an oriented graph. For the derivation of matrix model in this paper, however, we do not need to invoke the notion of lozenge tilings.} The coordinates of the $k$-th path at time $t$ (specified by the $t$-th dotted line) is given by an integer $h_k(t)$. 
It is easy to see that there is a one-to-one correspondence between the coordinates $h_k(t)$ satisfying the conditions above and the non-intersecting paths on the oriented graph. The inequality \eqref{eq.hnonintersecting} is translated into the non-intersecting condition. The step condition
\eqref{eq.hevolve} corresponds to the fact that we have two arrows for each vertex on the oriented graph, one with the same coordinate and another with coordinate increasing/decreasing by one unit.
Thus, the BPS partition function can be expressed as a sum over
non-intersecting paths, 
\begin{equation}
\begin{split}
Z_{\rm BPS}(q)& = 
\sum_{ \{h_i(t) \} \textrm{ : non-intersecting} } 
\prod_{t} q^{\sum_i h_i(t)}
\end{split},
\label{eq.C3sumNIP}
\end{equation}
where paths are assumed to satisfy the boundary condition \eqref{eq.hboundary}.

By the LGV formula (appendix \ref{sec.LGV}), \eqref{eq.C3sumNIP} is equivalent to
\begin{equation}
\begin{split}
Z_{\rm matrix}(q)& = 
\int \prod_{t} dh(t)~
%1/\Delta(X_{\tmax}) 
%\sum_{ \{h_i(t) \} } 
%
%q^{\Tr h_i(t)} \\
%
 \textrm{det}_{i,j} \left(
G(i,j;t)
\right),
\end{split}
\end{equation}
where the Green function $G(i,j;t)$ is given by
\beq
\begin{split}
& G(i,j;t) = q^{\sum_i h_i(t)} \\
& \quad \quad \times 
\left[ \delta\left(h_i(t+1)-h_j(1)\right)
+ \delta\left(h_i(t+1)-h_j(t)+\sigma(t+1/2)\right)
\right].
\end{split}
\eeq
The discrete coordinates $h_i(t)$ are turned into continuous variables. 
The delta functions enforce the condition \eqref{eq.hevolve}. 
The contributions including off-diagonal components of Green functions 
correspond to intersecting paths, which cancel out by the sign of the determinant.
We also need to set the boundary condition \eqref{eq.hboundary}
\beq
h_i(t)\in \{1,\ldots, N \}, \quad |t|\gg 1,
\label{eq.Xboundary}
\eeq
where $N$ is an integer which we take to infinity at the end of the computation. 
Keeping $N$ finite corresponds to taking only the first $N$ paths. 

The delta functions can be generated by introducing Lagrange multipliers
$\phi(t') ~  (t'\in \bZ+\frac{1}{2})$,
\beq
\begin{split}
 \frac{1}{N!} & \int d\phi(t')~ e^{-\Tr~ V_{t'}(\phi(t')) }~ \det(e^{i h_i(t'+\frac{1}{2}) \phi_j(t')}) \det(e^{-i h_i(t'-\frac{1}{2}) \phi_j(t')}) \\
&~=\textrm{det} \left[
 \delta(h_i(t'+\frac{1}{2})-h_j(t'-\frac{1}{2}))
+ \delta(h_i(t'+\frac{1}{2})-h_j(t'-\frac{1}{2})+\sigma(t'))
\right],
\end{split}
\nonumber
\eeq
where the potentials $V_{t'}(\phi(t'))$ depend on the signs 
of $t'$ and are given by
\beq
e^{-V_{t'}(\phi)}=
1 +e^{i \phi \sigma(t')}.
\eeq
We can enforce the boundary condition (\ref{eq.Xboundary}) by
limiting $-M \leq t \leq M$ for the height function $h(t)$ and 
$-M -\frac{1}{2} \leq t' \leq M + \frac{1}{2}$ for the
Lagrange multiplier $\phi(t')$ for sufficiently large $M$, and by 
introducing the factors,
\be
\Delta \left( e^{-i  \phi(-M-\frac{1}{2})}\right)
\Delta \left( e^{i  \phi(M+\frac{1}{2})}\right),
\label{VacuumWaveFunction}
\ee
at the initial and final points,
where $\Delta(e^{i\phi})$ is the Vandermonde determinant,
$$\Delta(e^{i\phi}) = {\rm det}_{kl}\left( e^{i k\phi_l}\right).$$
In the free fermion formalism in the previous subsection, the
two determinants in (\ref{VacuumWaveFunction})
correspond to the bra and ket states for the Fock vacuum. 
We also set the potentials at the two end points to vanish,
$V_{t'=-M-\frac{1}{2}}=V_{t'=M+\frac{1}{2}}=0$.
The partition function then takes the form\footnote{We drop 
an overall constant here for simplicity. In the final 
expression, we will present
the correct formula including the overall factor.}
\begin{equation}
\begin{split}
Z_{\rm matrix}(q)=  &\int 
\prod_t dh(t)   
\int \prod_{t'} d\phi(t') ~~
\Delta \left( e^{-i  \phi(-M-\frac{1}{2})}\right)
\Delta \left( e^{i  \phi(M+\frac{1}{2})}\right),
\\&
\times
 \prod_t  q^{\Tr\, h(t)}
\prod_{t'} e^{-\Tr\, V_{t'}(\phi(t'))} \det(e^{i h(t'+\frac{1}{2})\phi(t')}) 
\det(e^{- i h(t'-\frac{1}{2})\phi(t')}).
\end{split}
\end{equation}

We can turn this into a matrix integral. We use
the Itzykson-Zuber integral over
unitary matrix $U$ \cite{Harish-Chandra,Itzykson-Zuber}
\beq
\det(e^{i X_i Y_j})=
\Delta(X) \Delta(Y)
\int dU~ e^{i \Tr XUYU^{\dagger}},
\eeq
to generate squares of the Vandermonde determinants
$\Delta(h(t))^2$ and $\Delta(\phi(t'))^2$ for all $t$ and $t'$, 
except for $\phi(-M-\frac{1}{2})$ and $\phi(M+\frac{1}{2})$, for which
$\Delta(\phi(-M-\frac{1}{2}))$ and $\Delta(\phi(M+\frac{1}{2}))$
are generated.
The resulting expression can now be written as a matrix integral
\begin{equation}
\begin{split}
Z_{\rm matrix}(q)= &\int
 \prod_{t} dH(t)  
 \int \prod_{t'} d\Phi(t') 
~~ \frac{\Delta \left( e^{-i  \phi(-M-\frac{1}{2})}\right)}
{\Delta(\phi(-M-\frac{1}{2}))}
\frac{\Delta \left( e^{i  \phi(M+\frac{1}{2})}\right)}
{\Delta(\phi(M+\frac{1}{2}))}
\\&
\times 
%\\
%& 
\prod_{t} q^{\Tr~ H(t)}
\prod_{t'} e^{-\Tr~ V_{t'}(\Phi(t'))} e^{\Tr~ i \Phi(t') (H(t'+\frac{1}{2})
-H(t'-\frac{1}{2}))},
\label{eq.C3Zmultimatrix}
\end{split}
\end{equation}
where $H(t)$ ($t\in \mathbb{Z}$) and $\Phi(t')$ ($t'\in \mathbb{Z}+\frac{1}{2}$) are $N\times N$ 
Hermitian matrices whose eigenvalues are $(h_i(t))_{i=1,...,N}$ and $(\phi_i(t'))_{i=1,...,N}$, respectively.

%------------------------------------
\bigskip

The matrix model \eqref{eq.C3Zmultimatrix} is a multi-matrix model. However, $H(t)$ appear only linearly in the integrand, and can be trivially integrated out, yielding the constraints
$$
\Phi(t')=\Phi(t'-1)+ \log q\cdot{\bf 1}_{N\times N}=\Phi\left(-\frac{1}{2}\right)+ \left(t'+\frac{1}{2}\right) \log q\cdot {\bf 1}_{N\times N}.
$$
The matrix model then simplifies to a one-matrix.
The measure factor for $\Phi = \Phi(-\frac{1}{2})$ is
$$ d\Phi\cdot \left| \frac{\Delta(e^{i\phi})}{\Delta(\phi)}\right|^2
 = dU,$$
where $U=e^{i\Phi}$.  
The factor (\ref{VacuumWaveFunction}) we inserted to impose
the boundary condition turns the hermitian measure for $\Phi$
into the unitary measure for $U$. 
Thus, the matrix integral can be written as 
\beq
Z_{\rm matrix}= \int dU \ \Theta(U|q).
\label{eq.renomV}
\eeq
This gives another derivation of \eqref{CthreeMatrix}.

\subsubsection{Conifold and More General Geometries}

\begin{figure}[htbp]
\centering{\includegraphics[scale=0.35]{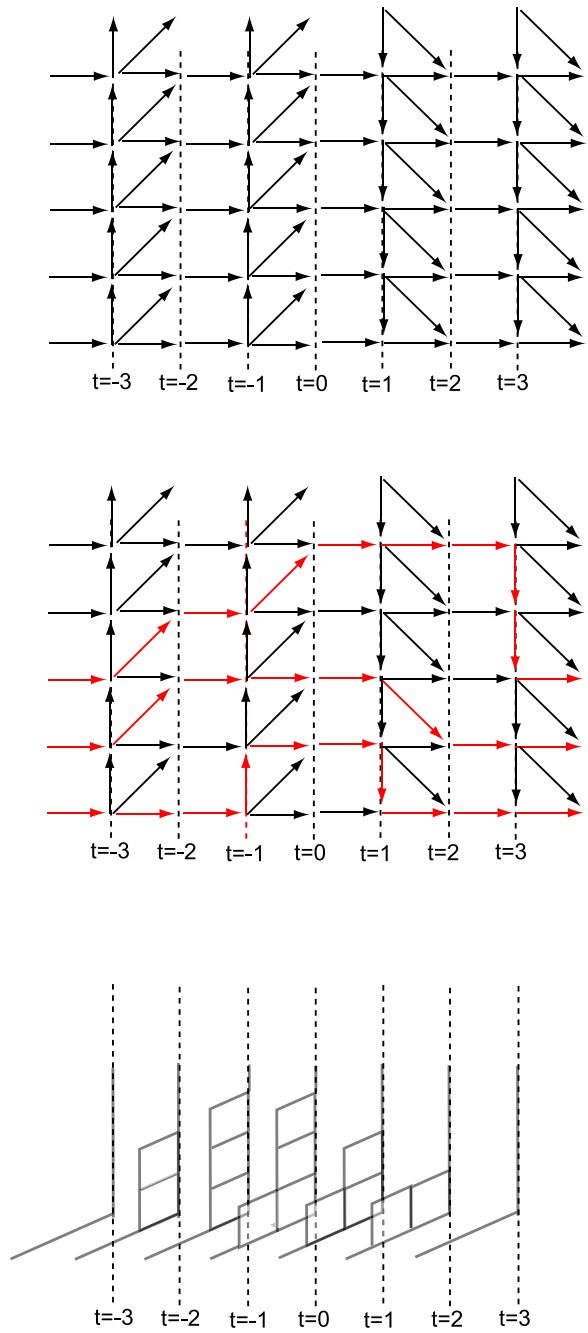}}
\begin{quote}
\caption{\emph{Top:
an oriented graph for the conifold.
Middle: an example of 3 non-intersecting paths on the graph shown
in red. Bottom: the corresponding evolution of Young diagrams.}}
\label{fig.conifoldorientedgraph}
\end{quote}
\end{figure}

Let us next describe the conifold in the non-commutative chamber. Again, we define $h_k(t)$ by \eqref{eq.hdef}.
We then have the non-intersecting condition \eqref{eq.hnonintersecting} and the boundary condition \eqref{eq.hboundary}. We also have, from \eqref{eq.lambdaconifoldinterlacing},
\begin{enumerate}
\item When $t$ is odd, 
\beq
h_k(t+1)-h_k(t)=0  ~ \textrm{or}~ \sigma(t+\frac{1}{2}).
\label{eq.hevolve0}
\eeq

\item When $t$ is even, 
\beq
\ldots \leq h_{k-1}(t+1) <h_k(t)\leq h_k(t+1) < h_{k+1}(t) \leq \ldots .
\label{eq.hevolve1}
\eeq
for $t\ge 0$ and 
\beq
\ldots \leq h_{k-1}(t) <h_k(t+1)\leq h_k(t) < h_{k+1}(t+1) \leq \ldots .
\label{eq.hevolve2}
\eeq
for $t<0$.
\end{enumerate}
The set of coordinates $h_k(t)$ satisfying the conditions above can be expressed as coordinates of non-intersecting paths of the oriented graph shown in figure \ref{fig.conifoldorientedgraph}. Let us show that this is indeed the case. When $t$ is odd, the story is similar to the $\bC^3$ example; we have two possibilities in \eqref{eq.hevolve0}, which corresponds to the two arrows starting from a vertex of the oriented graph, one with the same height and another with increasing/decreasing height by one unit. The situation changes when $t$ is even; during one time unit, after going one unit horizontally we can choose to go vertically as much as we want, as long as we respect the non-intersecting condition. In other words, the intersection of the $k$-th path and the time slice $t$ is an oriented interval, and when take $h_k(t)$ to be the value at the endpoint of the oriented interval, the interval is expressed as $[h_k(t), h_k(t+1)]$ (this is for $t<0$; for $t\ge 0$, we have $[h_k(t+1), h_k(t)]$ instead).
This means that the conditions \eqref{eq.hevolve1}, \eqref{eq.hevolve2} are translated into the non-intersecting conditions for paths. In the vertex operator formalism explain in section \eqref{ssec-MatrixFermions}, $t$ odd ($t$ even) corresponds to $\Gamma_{\pm}$ ($\Gamma'_{\pm}$).

\bigskip

The procedure to obtain the matrix model is similar to the $\bC^3$ example. The multi-matrix model is given as follows:
\begin{equation}
\begin{split}
Z_{\rm matrix}(q)=&
\int
\prod_{t} dH(t)
  \int \prod_{t'} d\Phi(t') 
~~ \frac{\Delta \left( e^{-i  \phi(-M-\frac{1}{2})}\right)}
{\Delta(\phi(-M-\frac{1}{2}))}
\frac{\Delta \left( e^{i  \phi(M+\frac{1}{2})}\right)}
{\Delta(\phi(M+\frac{1}{2}))}
\\
&\times \prod_{t}  q_t^{H(t)}
\prod_{t'} e^{-\Tr~ V_{t'}(\Phi(t'))}~~ e^{\Tr~ i \Phi(t') (H(t'+\frac{1}{2})
-H(t'-\frac{1}{2}))},\label{eq.conifoldZmatrix}
\end{split}
\end{equation}
The main differences from the $\bC^3$ example are that we have two parameters which depend on time $t$ as
\beq
q_t=\begin{cases}
q_0 & (t{\rm :odd}),\\
q_1 & (t{\rm :even}), 
\end{cases}
\eeq
and that the potential takes the form
\beq
e^{-V_{t'}(\phi)}=
\begin{cases}
1 +e^{i \phi \sigma(t')} & (t'{\rm :odd}), \\
1/(1 -e^{i \phi \sigma(t')}) & (t'{\rm :even}). \\
\end{cases}
\eeq
When $t$ is odd, there are 2 possibilities for $h_k(t+1)-h_k(t)$, which is the reason for the 2 terms in the potential. When $t$ is even, $e^{-V_{t'}(\phi)}=1+e^{-i \phi \sigma(t')}+e^{-2i \phi \sigma(t')}+\ldots$; this reflects the condition (which is part of \eqref{eq.hevolve2})
\begin{equation*}
h_k(t+1)=h_k(t)+m \sigma(t+1/2),~~ m=1, 2, 3, ...,
\end{equation*}
$i.e.$, $h_k(t+1)\ge h_k(t)$. The remaining conditions of \eqref{eq.hevolve2} is taken care of by the non-intersecting condition. 

Again, we can integrate out $M_t$'s, and the matrix model simplifies.
 When we diagonalize it, we finally have 
\beq
Z_{\rm matrix}=\int dU \det \frac{\Theta(U|q)}{\Theta(QU|q)},
\eeq
where $U \equiv e^{i\Phi(-\frac{1}{2})}$, 
$q\equiv q_0 q_1$ and $Q=-q_1$. 
The sign in $Q$ comes from the localization \cite{MR}, 
which should properly be taken into account in the definition of the matrix model.

\begin{figure}[htbp]
\centering{\includegraphics[scale=0.4]{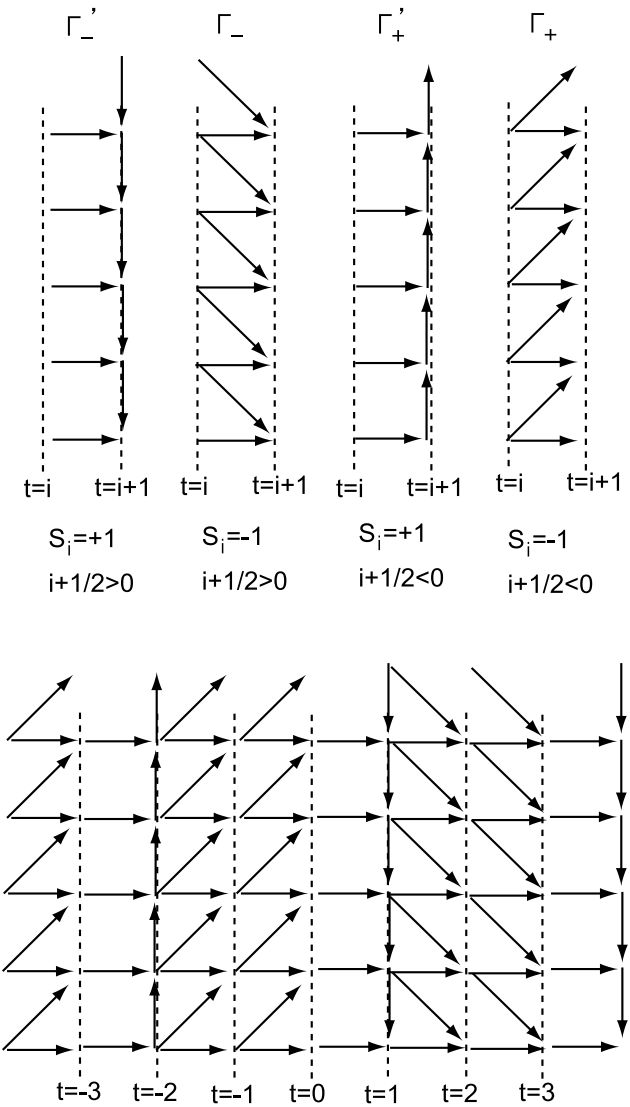}}
\begin{quote}
\caption{\emph{Top: An oriented graph in general cases are constructed by combining the 4 basic types of graphs shown here.
Bottom: An oriented graph for SPP, with $S_1=+1, S_2=-1, S_3=-1$.}}
\label{fig.SPPorientedgraph}
\end{quote}
\end{figure}

We can also repeat the same analysis for more general geometries. In particular, in the non-commutative chamber we obtain the results presented in section \ref{subsec.generalmatrix}. The oriented graph can be constructed from the data of $S_i$, by combining the 4 basic patterns (correponding to $\Gamma_+,\Gamma_-,\Gamma'_+, \Gamma'_-$) for each $i\in \bZ$; see figure \ref{fig.SPPorientedgraph}. As an example, the oriented graph for SPP in the noncommutative chamber with $S_1=+1, S_2=-1, S_3=-1$ is given in figure \ref{fig.SPPorientedgraph}. We stress that this approach is equivalent to the fermionic picture described earlier. 
For example, the sum over all possible paths in the region $t<0$ (or $t>0$) is encoded in the state $\langle\Omega_+|$ (respectively $|\Omega_-\rangle$). These states live in the Fock space associated to $t=0$, and can be expressed in terms of a sum over two-dimensional partitions from both fermionic and non-intersecting paths viewpoints. The correlator (\ref{ZOmega}) represents gluing paths extending in the $t<0$ region with paths in the $t>0$ region in a consistent way. Since the evolution rules of Young diagrams in more general chambers are already given in \cite{2009pyramid,NagaoVO,NY}, it is in principle straightforward to generalize the analysis to more general chambers. 
%Systematic investigation will be given elsewhere.

%%%%%%%%%%%%%%%%%%%%%%%%%%%%%%%%%%%%%%%%%%%%%%%%%%%%%%%%%%%%%%%%%%%%%%%%%%%%
\section{Discussion}

In this paper, we derived unitary matrix models of infinite-size matrices, which give the
counting of BPS bound states of D0 and D2-branes bound to  
a single D6-brane wrapping a toric Calabi-Yau manifold
$X$ without compact 4-cycle. These matrix models depend on 
a set of parameters $Q$, which keep track of the BPS charges, 
and the chamber parameters $n$. Both $Q$ and $n$ are 
associated to the K\"ahler moduli space ${\cal M}(X)$
of $X$. It turned out that these matrix models define
the topological string on another Calabi-Yau
manifold $Y$, whose moduli space contains two copies of ${\cal M}(X)$.
The parameters $Q$ and $n$ are unified as the K\"ahler moduli of $Y$.  
In addition, when the 't Hooft coupling $g_s N$ is finite, we found
yet more general manifold $\widetilde{Y}$. In the crystal model this finite 't Hooft coupling
has an interpretation of restricting a crystal configuration by a wall 
located at position $N$, and then the limit $N\to\infty$ 
provides mathematically rigorous definition of our models.

The relation between the BPS counting on $X$
and the topological string on $Y$ is clearest in the commutative and
the non-commutative chambers. In other chambers, there is a non-trivial
prefactor in the relation between the BPS partition function and 
the matrix model partition function. We hope to understand the origin
and the nature of the prefactor better. 

Our methods provide a rigorous derivation of
matrix models and spectral curves, which encode the mirror map 
expected from the remodeling conjecture \cite{BKMP}.
In this context it is interesting to note the subtlety related 
to the counting of MacMahon factors. For example, in the conifold 
example in the commutative chamber with either $Q=0$ or $e^{-T}=0$, 
we have one power 
of MacMahon function $M(q)$, which agrees with topological string 
result and Chern-Simons partition function.
However there is a mismatch by $M(q)^{1/2}$ between our matrix model 
integral formula and the topological string
partition function for SPP. Similar mismatches arise in matrix models 
derived in \cite{eynard-planch,SW-matrix,EynardTopological}.

The notion of the spectral curve also exists in the dimer model. In \cite{OY2},
 which discusses the thermodynamic limit of the crystal melting model, it was proven using the results of \cite{kos}, that genus 0 contribution of the DT partition function in the noncommutative chamber agrees with the genus 0 part of the topological string on the spectral curve of the dimer model, which is the mirror of $X$. 
An interesting problem is to understand how the spectral curve of 
the matrix model is related to that of the dimer model.

The holomorphic anomaly equations of topological string amplitudes can be
interpreted as the manifestation of their background independence 
\cite{BCOV,Witten}. The relation between the BPS partition function on $X$
and the topological string on $Y$ suggests that the wall crossing phenomenon
on $X$ may be related to the background independence on $Y$. 
In this context it would also be interesting to relate our analysis directly to 
the continuous limit of Kontsevich-Soibelman equations \cite{KS}.

In this paper we considered bound states of D6-D2-D0 branes. This analysis can be extended,
both from M-theory and matrix model, to include an additional D4-brane and associated 
open BPS invariants  \cite{AY,2010-openWalls}. Refined versions of our results can also be found 
using similar techniques \cite{2010-refined}.

%It would also be interesting to extend the results of this paper to the refined case; this would 
%presumably give rise to matrix models for $\beta$-ensembles, similarly as in 
%\cite{2009betaMatrix}.

\newpage

{\center \section*{Acknowledgments}}

We thank Mina Aganagic, Vincento Bouchard, Kentaro Hori, and
Yan Soibelman for discussions. 
H.~O. and P.~S. thank Hermann Nicolai and the Max-Planck-Institut
f\"ur Gravitationsphysik for hospitality. 
 
Our work is supported in part by the DOE grant DE-FG03-92-ER40701.
H.~O. and M.~Y. are also supported in part by  
the World Premier International Research Center Initiative of 
MEXT. H.~O. is supported in part by
JSPS Grant-in-Aid for Scientific Research (C) 20540256
and by the Humboldt Research Award.
P.~S. acknowledges the support of the European Commission under the Marie-Curie
International Outgoing Fellowship Programme and the Foundation for Polish
Science. M.~Y. is supported in part by the JSPS Research Fellowship for Young
Scientists and the Global COE Program for Physical Science 
Frontier at the University of Tokyo.

%\newpage 

\bigskip

\bigskip

\appendix

\section{Unitary Measure and the Migdal Integral}    \label{app-matrix}

Matrix models derived in this paper, either from fermionic or non-intersecting paths viewpoint, are of the form
$$
Z_{{\rm matrix}} = \int  dU e^{-\frac{1}{g_s} \textrm{Tr}\, V_{\rm unitary}(U)},
$$
where the unitary measure, after diagonalization $U = {\rm diag}(u_1, ..., u_n)$ with eigenvalues $u_i = e^{i\phi_i}$,
takes form
$$
  dU = \prod_k d\phi_k \ \prod_{i<j} (e^{i\phi_i} - e^{i\phi_j})
              (e^{-i\phi_i} - e^{-i\phi_j}) .
$$
This measure can be turned into the form involving the standard Vandermonde determinant $dU\to \prod_k du_k \prod_{i<j}(u_i-u_j)^2$ at the expense of introducing an additional term $T\log U$ to the matrix potential 
\be
V_{\rm unitary}(U)\to  V(U) = V_{\rm unitary}(U)+T\log U,\qquad \qquad T=g_s N.   \label{measure-T}
\ee
To find the resolvent for compact domain of eigenvalue distribution, arising from the initial unitary matrix ensemble, 
one can use results of \cite{mandal}. Namely, the resolvent $\omega(u)$ of the resulting matrix model can be solved using the Migdal integral, as also explained in \cite{Marino} and confirmed in explicit computations e.g. in \cite{SW-matrix,CGMPS}. 
In case of the one-cut matrix model this integral takes form
\be
\omega(u) = \frac{1}{2T} \oint \frac{dz}{2\pi i} \frac{\partial_z V(z)}{u-z}\frac{\sqrt{(u-a_+)(u-a_-)}}{\sqrt{(z-a_+)(z-a_-)}},  \label{Migdal}
\ee
where the integration contour encircles counter-clockwise two endpoints of the cut $a_{\pm}$. 
%The resolvent found in this form must satisfy the condition
%\be
%\lim_{p\to\infty} \omega(p) = \frac{1}{p}.    \label{Migdal-norm}
%\ee

In computing such Migdal integrals we often come across the situation where the derivative of the potential $\partial_z V(z)$ contains terms of the form $\frac{\log (z+c)}{z}$. In this case we find
\bea
\widetilde{\omega}_c(u) & = &  \frac{1}{2T} \oint \frac{dz}{2\pi i} \frac{\log (z+c)}{z(u-z)}\frac{\sqrt{(u-a_{+})(u-a_{-})}}{\sqrt{(z-a_{+})(z-a_{-})}} =                  \label{Migdal-log}  \\
& = & -\frac{1}{2 u T}\log \Big(\frac{\sqrt{(a_{+}+c)(a_{-}-u)} - \sqrt{(a_{-}+c)(a_{+}-u)}}{(u+c)(\sqrt{a_{-}-u}-\sqrt{a_{+}-u})}   \Big)^2   +    \nonumber \\
& & -\frac{\sqrt{(u-a_{+})(u-a_{-})}}{2 u T \sqrt{a_{+} a_-}} \log\Big( \frac{\sqrt{(a_{+}+c)a_{-}}-\sqrt{(a_{-}+c)a_{+}}}{c(\sqrt{a_+}-\sqrt{a_-})}  \Big)^2.   \nonumber
\eea
This result arises from contour integrals around poles at $z=0$ and $z=u$, as well as along the branch cut of the logarithm $(-\infty,-c)$. To find the latter contributions the following integral is useful
%\footnote{Note a typo in \cite{CGMPS}.}
$$
\int\frac{dx}{(x-u)\sqrt{(x-a)(x-b)}} = -\frac{1}{\sqrt{(u-a)(u-b)}} \log \frac{(\sqrt{(x-a)(b-u)} - \sqrt{(x-b)(a-u)})^2}{(u-x)\sqrt{(u-a)(u-b)}}.
$$

\bigskip

In particular, for the conifold matrix model with the potential given in (\ref{Vconifold}), the resolvent can be expressed as
\be
\omega(u) = \widetilde{\omega}_{Qe^{\tau}}(u) -\widetilde{\omega}_Q(u) + \frac{T - \log(Qe^{\tau})}{2T}\Big( \frac{\sqrt{(u-a_+)(u-a_-)}}{u\sqrt{a_{+} a_{-}}} + \frac{1}{u} \Big).   \label{omega-coni-start}
\ee
In consequence we find that the resolvent is given by a sum of two terms, which in the limit $u\to\infty$ are respectively constant and of order $1/u$. Imposing the asymptotic condition on the resolvent $\omega(u)\sim 1/u$ given in (\ref{eq.omegaboundary}) implies that the constant term must vanish, while $\sim 1/u$ term must have a proper coefficient. This leads to the result (\ref{conifoldresolvent}), and
moreover gives rise to the two equations  (\ref{ab-eq1}) and (\ref{ab-eq2}) for the endpoints of the cut $a_{\pm}$. The solution to these equations is given in (\ref{aAll}). For various computations concerning this conifold example it is advantageous to use the identifies
\bea
a_{+} a_{-} & = & \Big(  \frac{1-Q\epsilon^2}{1-\mu\epsilon^2} \Big)^2,  \nonumber \\
(a_{+}+Q)(a_{-}+Q) & = & \Big(  \frac{1-Q(1-\epsilon^2+\mu\epsilon^2)}{1-\mu\epsilon^2} \Big)^2,  \nonumber \\
(1+a_{+}\mu)(1+a_{-}\mu) & = & \Big(  \frac{1-\mu(1-\epsilon^2+Q\epsilon^2)}{1-\mu\epsilon^2} \Big)^2 .  \nonumber 
\eea

%

%*******************************************************************

\section{Free Fermion Formalism}   \label{app-fermion}

For completeness we review free fermion formalism \cite{jimbo-miwa} following conventions of \cite{2009pyramid,YoungBryan}. We start with the Heisenberg algebra
$$
[\alpha_m,\alpha_{-n}] = n \delta_{m,n}
$$
and define
$$
\G_{\pm}(x) = e^{\sum_{n>0} \frac{x^n}{n}\alpha_{\pm n}}, \qquad \qquad \G'_{\pm}(x) = e^{\sum_{n>0} \frac{(-1)^{n-1}x^n}{n}\alpha_{\pm n}}.
$$
They act on fermionic states $|\mu\rangle$ corresponding to partitions $\mu$ as 
\bea
\G_-(x) |\mu\rangle  =  \sum_{\lambda \pg \mu} x^{|\lambda|-|\mu|}|\lambda\rangle,  & &\qquad \qquad  
\G_+(x) |\mu\rangle  =  \sum_{\mu \pg \lambda} x^{|\mu|-|\lambda|}|\lambda\rangle,      \\
\G'_-(x) |\mu\rangle =  \sum_{\lambda \succ \mu} x^{|\lambda|-|\mu|}|\lambda\rangle, & & \qquad \qquad
\G'_+(x) |\mu\rangle  =  \sum_{\mu \succ \lambda} x^{|\mu|-|\lambda|}|\lambda\rangle,
\eea
where $\succ$ and $\pg$ are interlacing relations defined in (\ref{def-succ}) and (\ref{def-pg}). These operators satisfy commutation relations 
\bea
\G_+(x) \G_-(y) & = & \frac{1}{1-xy} \G_-(y) \G_+(x),   \\
\G'_+(x) \G'_-(y) & = & \frac{1}{1-xy} \G'_-(y) \G'_+(x),   \\
\G'_+(x) \G_-(y) & = & (1+xy) \G_-(y) \G'_+(x),   \\
\G_+(x) \G'_-(y) & = & (1+xy) \G'_-(y) \G_+(x).
\eea

We also introduce various colors $q_g$ and the corresponding operators $\widehat{Q}_g$
$$
\widehat{Q}_g|\lambda\rangle = q_g^{|\lambda|}|\lambda\rangle.
$$

They commute with $\G$ operators as
\bea
\G_+(x) \widehat{Q}_g = \widehat{Q}_g \G_+(x q_g), & & \qquad \qquad
\G'_+(x) \widehat{Q}_g = \widehat{Q}_g \G'_+(x q_g), \\
\widehat{Q}_g \G_-(x) = \G_-(x q_g) \widehat{Q}_g, & & \qquad \qquad 
\widehat{Q}_g \G'_-(x) = \G'_-(x q_g) \widehat{Q}_g.
\eea

%%%%%%%%%%%%%%%%%%%%%%%%%%%%%%%%%%%%%%%%%%%%%%%%%%%%%%%%%%%%%%%%%%%%%%%%%%%%%%%

\section{LGV Formula}\label{sec.LGV}

In this appendix we explain the Linstr{\"o}m-Gessel-Viennot (LGV) formula \cite{Lindstrom,GesselV}, which is crucial for the derivation of the matrix model in section \ref{subsec.derivation2}.

Consider an oriented graph without closed loops. We assume that a weight $w(e)$ is assigned to each edge $e$ of the graph. We consider $N$ particles which follow paths $p_i$, each starting at vertices $a_i$ and ending at $b_i$ ($i=1,\ldots, N$). For such paths $P=\{p_i:a_i\to b_i\}$, we assign a weight
\beq
 w(p_i)=\prod_{e\in p_i} w(e).
\eeq

What we want to compute is the quantity
\beq
F(\{a_i\},\{b_i\})=\sum_{P \textrm{: non-intersecting}} \prod_i w(p_i),
\eeq
where the summation is over non-intersecting paths.
LGV formula states that this can be computed by summing over general (meaning, including intersecting) paths.
More precisely, when we define the ``Green function''
\beq
G(a_i,b_j)= \sum_{p \textrm{: a path from } a_i \textrm{ to } b_j} w(p),
\eeq
then LGV formula states that
\beq
F(\{a_i\},\{b_i\})=\det_{i,j}(G(a_i,b_j)).
\eeq
The proof is elementary, and proceeds by checking that contributions from intersecting paths cancel out due to the sign in the definition of the determinant. The determinant in the formula can be thought of as a discretized version of a Vandermonde determinant for free fermions, representing the Coulomb repulsions among particles.

Now consider a more general situation. Suppose that we are given a set of vertices $\{a_i(k)\}$, where $k=1,\ldots, L$. We consider $N$ particles, with the following condition: $i$-th particle starts from $a_i(0)$, goes through $a_i(1)$, then $a_i(2)$, \ldots, and finally arrives at $a_i(L)$. Then the multiplicative property of the determinant says that 
\beq
\det_{i,j}(G(a_i(1),a_j(L)))=\prod_{k=1}^{L-1}  \det_{i,j}(G(a_i(k),a_j(k+1)))
\eeq
This is the expression we need in the main text.

%%%%%%%%%%%%%%%%%%%%%%%%%%%%%%%%%%%%%%%%%%%%%%%%%%%%%%%%%%%%%%

\end{document}